\documentclass{article}

\usepackage{amsmath,amssymb,amsfonts}
\usepackage{array}
\usepackage{textcomp}
\usepackage[hyphens]{url}
\usepackage{graphicx}
\usepackage{cite}
\usepackage{booktabs}
\usepackage{tabularx}
\usepackage{moresize}
\usepackage{multirow}
\usepackage{hyperref}
\usepackage{wrapfig}
\usepackage{multicol}
\usepackage[a4paper, total={7in, 10in}]{geometry}
\usepackage{caption}
\usepackage{fancyhdr}

\newenvironment{singlecolelement}
    {\par\medskip\noindent\minipage{\linewidth}}
    {\endminipage\par\medskip}

\begin{document}

\title{Project-wise Comparison of Software Birthmarks Using Weighted Partial Similarity}

\author{
Nikolay Fedorov
\thanks{Corresponding author. N. Fedorov is with Graduate School of Environmental, Life, Natural Science and Technology, Okayama University, Okayama, Japan}
\\ \href{mailto:fedorov_n@s.okayama-u.ac.jp}{fedorov\_n@s.okayama-u.ac.jp}, 
\and
Akito Monden
\thanks{A. Monden is with Faculty of Environmental, Life, Natural Science and Technology, Okayama University, Okayama, Japan},
\\ \href{mailto:monden@okayama-u.ac.jp}{monden@okayama-u.ac.jp}, 
\and
Hiroki Inayoshi
\thanks{H. Inayoshi is with Faculty of Environmental, Life, Natural Science and Technology, Okayama University, Okayama, Japan},
\\ \href{mailto:inayoshi@okayama-u.ac.jp}{inayoshi@okayama-u.ac.jp}, 
\and
Haruaki Tamada
\thanks{H. Tamada is with Faculty of Information Science and Engineering, Kyoto Sangyo University, Kyoto, Japan},
\\ \href{mailto:tamada@cc.kyoto-su.ac.jp}{tamada@cc.kyoto-su.ac.jp}, 
\and
Masateru Tsunoda
\thanks{M. Tsunoda is with Faculty of Informatics, Cyber Informatics Research Institute, Kindai University, Higashiosaka-shi, Japan}
\\ \href{mailto:tsunoda@info.kindai.ac.jp}{tsunoda@info.kindai.ac.jp}, 
}

\maketitle

\thispagestyle{fancy}

\begin{multicols*}{2}
\begin{abstract}

Software birthmarks provide a robust approach to detecting code plagiarism even under substantial modifications, while distinguishing independently developed software.
Existing similarity measures are typically applied at the \textit{module level} (e.g., source or class files). 
However, in practice, software reuse often occurs at the \textit{project level}, where only a subset of modules may be reused. This setting introduces two key challenges: (1) \textit{partial reuse}, where reused modules constitute only a small fraction of the project, and (2) \textit{incidental similarity} from small modules, which can lead to false positives.

In this paper, we establish a framework for \textit{project-wise} birthmark comparison based on a symmetric aggregation of module-level similarities. 
On top of this framework, we propose two complementary mechanisms to address the above challenges.
First, we introduce a \textit{weighting scheme} that assigns higher importance to larger modules, reducing the influence of noisy matches from small modules.
Second, we propose a \textit{partial similarity} method that focuses on the top fraction of highly similar module pairs, enabling robust detection of partial reuse.

We evaluate the proposed approach on 35 open-source Java projects across ten categories, where different versions of the same project are treated as reuse cases.
The dataset and experimental artifacts are made publicly available to support reproducibility.
Performance is assessed using two complementary properties of software birthmarks, \textit{resilience} and \textit{credibility}, combined via their harmonic mean.
The results show that the proposed method consistently outperforms existing approaches, achieving robust and stable detection of partial code reuse at the project level.
\end{abstract}

\section{Introduction}
\label{sec:intro}

Software plagiarism and unauthorized code reuse have become significant concerns in modern software development, particularly with the widespread availability of open-source software (OSS) and code-sharing platforms~\cite{kim2014OSS, Lopes2017dejavu}.
These unethical practices involve copying or reusing code without proper attribution or permission.

Recent studies have highlighted the prevalence of unauthorized code reuse across various software development platforms.
Research on StackOverflow\footnote{\url{https://stackoverflow.com}} has examined how uploaded code snippets are reused without proper attribution~\cite{an2018stackoverflow,romansky2018sourcerer}, while other work has focused on detecting code cloning in open-source projects~\cite{kim2014OSS, duan2017identify}.
Golubev et al.~\cite{golubev2020codeBorrowingGithub} analyzed Java projects in the Public Git Archive~\cite{Markovtsev2018gitarchive} and reported that approximately 29.6\% of code fragments may have been incorporated without authorization, with about 9.4\% potentially violating original licenses.
Similarly, Lopes et al.~\cite{Lopes2017dejavu} found that nearly 70\% of code hosted on GitHub\footnote{\url{https://github.com/}} consists of clones of existing files.
These findings underscore the urgent need for reliable techniques to detect code reuse in real-world software systems.

Software birthmarks have been proposed as an effective technique for detecting software plagiarism~\cite{tamada2003detecting}, offering an advantage over software watermarking methods~\cite{zhu2005surveyWatermarking,monden2000watermarkingJava} in that they do not require any modification to the original files.
A birthmark captures intrinsic characteristics of a program, enabling similarity analysis even when the code has undergone transformations such as obfuscation~\cite{tamada2004design,tamada2005java}.
Various types of birthmarks have been proposed, including instruction-based~\cite{tamada2005java, myles2005Kgram}, structure-based~\cite{myles2004Detecting, chae2013graphBasedPlagiarismDetection}, and API call-based approaches~\cite{lee2016api, schuler2006api, Choi2009staticapi, wang2009behavior, schuler2007dynamic, wang2009detecting}.
Similarity functions such as cosine similarity and the Jaccard coefficient are commonly used to quantify similarity between modules~\cite{nazir2019birthmarkDesignEstimation, Tian2014MultithreadedBirthmark}.

Most existing studies focus on \textit{module-level} comparison (e.g., source or class files), which is effective for small-scale scenarios such as plagiarism detection in programming assignments~\cite{JoyPlagiarismInAssignments, LeetalEducating}.
However, this approach becomes insufficient in large-scale and real-world settings~\cite{liu2006gplag}, where software systems are composed of numerous modules and reuse may occur only partially.

Real-world cases further highlight this limitation.
For example, Vizio’s SmartCast platform allegedly violated the GNU GPL by reusing code from BusyBox and coreutils~\cite{sfconservancyVizioCaseStatus, sfconservancyComplaint}, and Teraproc’s OpenLava project was found to incorporate proprietary IBM code~\cite{IBMvTeraprocCase}.
Such cases demonstrate that plagiarism often occurs at the level of entire software systems rather than individual files.

These observations motivate a two-stage detection process:
(1) \textit{project-level retrieval}, which identifies candidate projects that may contain reused code, and
(2) \textit{module-level verification}, which performs detailed analysis of reuse.
Without an effective \textit{project-wise similarity} measure, the first stage cannot be performed reliably, making large-scale plagiarism detection impractical.

Despite its importance, project-wise birthmark comparison has received limited attention.
One reason is the lack of systematic evaluation settings.
Another challenge lies in aggregating module-level similarities into a reliable project-level measure.
This setting introduces two key difficulties:
(1) \textit{partial reuse}, where reused code appears only in a subset of modules, and 
(2) \textit{incidental similarity}, where small or generic modules produce misleading similarity across unrelated projects.

In this paper, we address challenges by establishing a method for \textit{project-wise} birthmark comparison.
Our approach builds upon module matching and similarity aggregation, and introduces two key mechanisms.
First, a \textit{weighting scheme} assigns higher importance to larger modules, which are more likely to contain meaningful reuse.
Second, a \textit{partial similarity} strategy focuses on the most relevant subset of module pairs, enabling robust detection even when reuse is limited.
These mechanisms jointly improve both resilience to modifications and credibility in distinguishing unrelated projects.

To enable systematic evaluation, we construct a dataset of 35 open-source Java projects across 10 categories, where different versions of the same project are treated as reuse cases.
In this study, we focus on static $k$-gram birthmarks to provide a controlled and consistent evaluation setting, while the proposed framework itself is general and applicable to other types of birthmarks.

Our contributions are summarized as follows:
\begin{itemize}
\item We formulate the problem of \textit{project-wise} birthmark comparison and position it as a key component in large-scale plagiarism detection.
\item We identify two fundamental challenges---partial reuse and incidental similarity---in project-level similarity measurement.
\item We propose a weighted and partial similarity framework to address these challenges.
\item We construct a benchmark dataset and evaluation methodology based on resilience and credibility. 
\item We demonstrate through extensive experiments that the proposed method improves detection performance over existing approaches.
\end{itemize}

The remainder of this paper is organized as follows.
Section~\ref{sec:birthmarks} introduces the concept of software birthmarks.
Section~\ref{sec:project-wise-sim} formulates the problem of project-wise similarity and discusses its inherent limitations.
Section~\ref{sec:proposed-framework} presents the proposed framework for project-wise birthmark comparison.
Section~\ref{sec:experiment} describes the experimental design, including the dataset and preprocessing methods.
Section~\ref{sec:results} reports the experimental results and provides detailed analysis.
Section~\ref{sec:validity-threats} discusses threats to validity, and Section~\ref{sec:conclusion} concludes the paper.

\section{Software Birthmarks}
\label{sec:birthmarks}
\subsection{Definitions}
\label{sec:birthmarks-definition}
\noindent 
A software birthmark is a representation of a program that captures its intrinsic characteristics and enables similarity-based comparison between programs~\cite{tamada2005java, tamada2004design, lee2016api, nazir2019birthmarkDesignEstimation}.

Formally, let $B(P)$ denote a function that extracts a birthmark from a program $P$.
If program $Q$ is an exact copy of $P$ (i.e., $Q \equiv P$), then their birthmarks are identical:
\begin{equation}
B(P) = B(Q).
\label{eq:birthmark-equal}
\end{equation}

In practice, software birthmarks are compared using similarity functions to detect unauthorized code reuse or plagiarism.
A detailed explanation on birthmark similarity calculation will be given in Section~\ref{sec:birthmark-sim-calc}.

Beyond plagiarism detection, they have also been applied to tasks such as malware analysis and bug detection~\cite{lee2016api, cesare2011malwareVariantDetection, Charland2012clonesearch, duan2017identify, jang2012redebug, Roy2007softwareclonedetection}.

A desirable software birthmark should satisfy two key properties~\cite{Alejandro2023autoextraction,lee2016api,yuan2018weightSequencesBirthmark}:

\begin{itemize}
\item \textbf{\textit{Credibility}}: If programs $P$ and $Q$ are independently developed, their birthmarks should not be similar:
\begin{equation}
B(P) \nsim B(Q),
\label{eq:credibility}
\end{equation}

\item \textbf{\textit{Resilience}}: If program $Q$ is derived from $P$, their birthmarks should remain similar:
\begin{equation}
B(P) \sim B(Q).
\label{eq:resilience}
\end{equation}
\end{itemize}

Traditionally, resilience has been defined with respect to \textit{semantic-preserving transformations}, such as code obfuscation or decompile--recompile processes~\cite{tamada2004dynamic,myles2005Kgram,Yadegari2015automaticdeobfuscation,collberg1997obfuscatingtransformations}.  
However, such transformations do not fully reflect realistic software reuse scenarios.
In practice, reused code is often modified, extended, or partially rewritten to fit new contexts, introducing functional changes beyond purely semantic-preserving transformations~\cite{Roy2007softwareclonedetection}. 

In this work, we adopt a broader interpretation of resilience that includes such practical modifications.
Specifically, we consider a program $Q$ to be derived from $P$ if $Q$ is created by reusing and modifying parts of $P$, even when additional functionality is introduced.
Under this interpretation, resilience captures the ability of a birthmark to preserve similarity when reused code is modified and integrated into a different software system.
In practical scenarios, an adversary may reuse code from an existing project, apply modifications such as extensions, refactoring, or partial rewriting, and distribute it as a new product~\cite{Roy2007softwareclonedetection}.
A resilient birthmark should still enable the detection of such derivation despite these changes.
In our evaluation, we approximate this scenario by comparing birthmarks across different versions of the same software, where later versions include modifications and extensions over earlier ones.

While resilience is essential for detecting reused code, credibility is equally important to avoid false positives~\cite{tamada2004dynamic,myles2005Kgram}.
In module-level comparison, achieving high credibility is inherently challenging, as small or generic modules may exhibit incidental similarity even when independently developed~\cite{Lim2009controlFlowEdges}. 
By ``module'' we refer to a source file/compiled source file, as well as any other project component, from which the birthmark is extracted.

These observations highlight a fundamental limitation of existing birthmark definitions, which are commonly centered on module-level comparison.
When moving to project-level analysis, where a program consists of many modules, both resilience and credibility must be reconsidered in an aggregated context.
In particular, project-level similarity should remain robust to partial reuse while suppressing incidental similarity arising from small modules.

\subsection{Birthmark Types Overview}
\label{sec:birthmark-types}
\noindent
Software birthmarks can be broadly characterized along two dimensions: (1) the analysis approach and (2) the program elements they capture.
With respect to the analysis approach, birthmarks are classified as either \textit{static} or \textit{dynamic}.
Static birthmarks are extracted without executing the program (e.g., from source code, bytecode, or binaries), whereas dynamic birthmarks are obtained by observing runtime behavior under specific inputs~\cite{lee2016api, nazir2019birthmarkDesignEstimation, lim2009controlFlowInformation, jhi2011programchar, tian2015plagiarismDetectionDynamicKey}.
Dynamic approaches tend to be more resilient to program transformations, but their effectiveness depends on input coverage.
In contrast, static approaches are easier to apply at scale and provide full code coverage, but may be more sensitive to modifications.

With respect to the program elements they capture, birthmarks are commonly categorized into three groups~\cite{lee2016api}:

\begin{itemize}
    \item
\textbf{Instruction-based birthmarks}, which represent low-level program elements such as instruction sequences (e.g., static k-grams of Java opcodes~\cite{myles2005Kgram} and dynamic k-grams~\cite{lu2007dynamicopcode, bai2008dynamic}).
They are simple and scalable, but can be sensitive to code modifications~\cite{lee2016api,myles2005Kgram}.

    \item 
\textbf{Structure-based birthmarks}, which capture structural relationships such as control flow birthmarks~\cite{lim2009controlFlowInformation}, whole path birthmarks~\cite{myles2004Detecting}, program dependence paths~\cite{liu2006gplag}, and project logic paths~\cite{zhang2014programlogic}.
They can represent higher-level program characteristics, but often incur higher computational cost and may be sensitive to compiler optimizations~\cite{tamada2007design, lee2016api}.

    \item  
\textbf{API-based birthmarks}, which are derived from the usage patterns of standard libraries or system calls.
For example, such as used classes (UC) and sequences of method calls (SMC)~\cite{tamada2005java}, static API call sequences~\cite{lee2016api}, dynamic API call sequences and frequencies~\cite{tamada2004dynamic, schuler2006api, tian2013dkisb}.  
They are generally more robust to program modifications, as API usage patterns are harder to alter without affecting functionality~\cite{tamada2007design, lee2016api}. 
\end{itemize}

These categories reflect inherent trade-offs between robustness, granularity, and computational cost.
Importantly, the challenges addressed in this paper---such as partial reuse and incidental similarity---are not specific to any particular type of birthmark, but arise from how birthmark similarity is aggregated at the project level.
Therefore, the proposed project-wise comparison framework is designed to be applicable across different types of birthmarks, specifically, birthmarks that are calculated and compared per module or per specific part of the project. 

\subsection{k-gram Birthmark}
\label{sec:k-gram-birthmark}
\noindent
In this study, we employ the static k-gram birthmark for experimental evaluation.
A k-gram birthmark represents a program as a set of contiguous instruction sequences of length $k$ (i.e., k-grams), extracted statically from its code~\cite{myles2005Kgram, lee2016api}.
Given a program $P$, its birthmark is defined as:
\begin{equation}
B(P) = \bigl\{\text{k-gram}_i \mid i = 1,\dots,K\bigr\},
\label{eq:k-gram_birthmark}
\end{equation}
where $K$ denotes the number of unique k-grams.
In this representation, the order and frequency of k-grams are typically ignored, which reduces sensitivity to minor code transformations~\cite{myles2005Kgram}.

The k-gram birthmark is widely used in prior studies due to its simplicity and scalability~\cite{myles2005Kgram, lu2007dynamicopcode, bai2008dynamic}.
It can be extracted from various program representations (e.g., source code, bytecode, or binaries) with minimal preprocessing, making it suitable for large-scale analysis.
Moreover, the parameter $k$ controls the granularity of the representation, enabling a trade-off between robustness and discriminative power.

While k-gram birthmarks are known to be sensitive to code modifications and may grow in size for large programs~\cite{lee2016api,yuan2018weightSequencesBirthmark}, these limitations are acceptable for the purpose of this study.
Our goal is not to optimize the birthmark itself, but to evaluate how similarity should be computed at the project level.
In this context, k-grams provide a simple and widely adopted baseline that allows us to isolate the effect of the proposed similarity framework.
In addition, for Java programs, publicly available tools such as \textit{pochi}\footnote{\url{https://github.com/tamada/pochi}} facilitate birthmark extraction, further lowering the barrier to use.

Importantly, the proposed project-wise similarity method is not specific to k-gram birthmarks and is equally applicable to other types of birthmarks, including those with stronger resilience to code transformations.

\subsection{Birthmark Similarity Calculation}
\label{sec:birthmark-sim-calc}
\noindent
To determine whether code has been reused between two software modules, it is necessary to quantify the similarity between their birthmarks.
This is typically achieved by a \textit{similarity function}, which maps a pair of birthmarks to a numeric value in $[0.0,1.0]$, where higher values indicate greater similarity.

Formally, for two modules $p$ and $q$, let $X = B(p)$ and $Y = B(q)$ denote their birthmarks.
A module-wise similarity function is defined as:
\begin{equation}
sim(p,q) = f(X,Y),
\end{equation}
where $f$ is a similarity function.

Application of various similarity functions have been proposed in prior work~\cite{Tian2014MultithreadedBirthmark,lee2016api, myles2005Kgram, tamada2004design}, which can be broadly categorized into three groups:
\begin{itemize}
\item \textbf{Vector-based similarity}, such as cosine similarity, where birthmarks are transformed into feature vectors and compared based on their geometric similarity.

\item \textbf{Set-based similarity}, such as the Jaccard coefficient, Dice index, and Simpson index, which measure the overlap between sets of birthmark elements.

\item \textbf{Sequence-based similarity}, such as edit distance, which compares ordered sequences of birthmarks and captures structural differences between them.
\end{itemize}

However, the described functions would be applicable only for instruction-based and API-based birthmarks, while structure-based would require different similarity computation approach, due to their usual representation as graphs~\cite{lee2016api}.  

These similarity functions exhibit different characteristics.
Vector-based methods capture global distributional patterns, set-based methods emphasize shared elements, and sequence-based methods preserve ordering information.
The choice of similarity function therefore affects the sensitivity to different types of code modifications.

In this study, we employ representative similarity functions from each category to ensure a comprehensive evaluation (the list of employed functions will be provided in Section~\ref{sec:module-wise-sim-functions}).
However, the focus of this paper is not on the choice of module-level similarity function itself, but on how such similarities should be aggregated at the project level.

Finally, it is important to note that module-wise similarity alone is insufficient for large-scale plagiarism detection.
While it can identify reuse between individual modules, it does not directly provide a mechanism to compare entire software projects.
This limitation motivates the need for project-wise similarity measures, which will be discussed in the next section.

\section{Project-wise Similarity: Problem and Limitations} 
\label{sec:project-wise-sim}
\subsection{Problem Formulation}
\label{sec:problem-formulation}
\noindent
In large-scale software ecosystems, a vast number of projects exist, each consisting of many modules (e.g., source files, classes, etc.).
In this context, detecting code reuse cannot rely solely on module-wise comparison.

Let a project $P$ be represented as a set of modules:
\begin{equation}
P = \{p_1, p_2, \dots, p_n\},
\end{equation}
where each module $p_i$ is associated with a birthmark $B(p_i)$.
Given two projects $P$ and $Q$, module-wise similarity can be computed for all pairs:
\begin{equation}
\{sim(p_i, q_j)\}, \quad \forall p_i \in P, \forall q_j \in Q.
\end{equation}

However, this formulation raises several fundamental challenges.

First, module-wise comparison produces a large number of similarity scores across module pairs, but does not provide a clear mechanism for interpreting these results at the project level.
Even if all pairwise similarities are computed, it remains unclear how to determine whether two projects are related as a whole.

Second, in realistic scenarios, a developer or analyst is typically given a target project $P$ and must search for potentially related projects $Q$ among a vast collection of candidates.
Without a project-level similarity measure, there is no principled way to rank or filter candidate projects for further inspection.

Third, software reuse is often \textit{partial}.
A project may reuse only a subset of modules from another project, while many other modules remain unrelated.
This dilutes the signal of reuse when considering all module pairs uniformly.

Finally, small or generic modules may exhibit incidental similarity, leading to false positives~\cite{Lim2009controlFlowEdges}.
Such noise can dominate the similarity signal if all modules are treated equally.

These challenges indicate that module-wise similarity is insufficient as a standalone approach.
Instead, it is necessary to define a project-wise similarity function:
\begin{equation}
Sim(P,Q) = F\bigl(\{sim(p_i,q_j)\}\bigr),
\end{equation}
which aggregates module-wise similarities into a single project-level score, enabling direct comparison of projects.

The design of such a function must address the following requirements:
\begin{itemize}
\item Robustness to partial reuse, where only a subset of modules is reused;
\item Robustness to incidental similarity arising from small or generic modules;
\item The ability to provide meaningful project-level scores that support ranking and filtering of candidate projects in large-scale software ecosystems.
\end{itemize}

However, as we discuss next, existing approaches do not fully satisfy these requirements.

\subsection{Limitations of Existing Project-wise Similarity Methods}
\label{sec:limits-of-existing-proj-sim-methods}
\noindent
Several approaches have been proposed to compute similarity between software projects.
In this section, we review representative methods and discuss their limitations with respect to the requirements identified in Section~\ref{sec:problem-formulation}.

\vspace{2mm}
\paragraph{Baseline method (Lee et al.)}
Lee et al.~\cite{lee2016api} proposed a project similarity measure based on procedure-wise (due to the work focusing on the API-based birthmark) similarities.
We adapt this method by using modules in place of procedures. 
For each module in project $P$, the maximum similarity to modules in $Q$ is computed, and the results are aggregated as follows:
\begin{equation}
\label{eq:baseline_method}
    Sim_p(P,Q) =
    \frac{
        2 
        \times 
        \sum_{i=1}^{n} 
        \max 
        \bigl( 
            \{sim(p_i,q_j) \mid j=1,\dots,m\} 
        \bigr)
    }{
        n + m
    },
\end{equation}
where $n$ and $m$ are the number of modules in $P$ and $Q$ respectively.

This method is simple but not symmetric, i.e., $Sim_p(P,Q) \neq Sim_p(Q,P)$, which leads to inconsistent results depending on the order of the inputs.

\vspace{2mm}
\paragraph{Aggregated similarity}
To address this limitation, our previous work~\cite{fedorov2024comparison} introduced \textit{aggregated similarity}, which computes a symmetric project-level score by considering high-similarity module pairs from both projects.

Formally, let $TopN(p_i, Q)$ denote the set of the top-$N$ highest similarity values between module $p_i$ and all modules in $Q$.
Aggregated similarity is defined as:
\begin{equation}
Sim_{agg}(P,Q) =
\frac{
\sum U
}{
|U|
},
\end{equation}
where $U$ is the union of all selected similarity values:
\begin{equation}
    U =\{TopN(p_i, Q)\} \cup \{TopN(q_j, P)\},\quad \forall p_i \in P, \forall q_j \in Q
\end{equation}

By incorporating the parameter $N$, this method considers multiple high-similarity matches per module, which, in theory, would help to reduce the impact of spurious matches, providing a more stable aggregation, compared to using only a single maximum, while also addressing cases where original module is plagiarized by several modules (i.e., parts of the original being split into many modules to avoid detection).
However, the previous study was conducted only for $N = 1$, therefore, the effectiveness of higher values of $N$ is yet to be tested~\cite{fedorov2024comparison}.

On the other hand, despite these improvements, aggregated similarity still treats all selected module pairs uniformly.
As a result, it does not explicitly account for differences in module importance or size, and remains sensitive to incidental similarity arising from small or generic modules.
Furthermore, it does not explicitly address partial reuse, where only a subset of modules should dominate the similarity score.

\vspace{2mm}
\paragraph{Summary of limitations}
Overall, existing project-wise similarity methods do not fully satisfy the requirements identified in Section~\ref{sec:problem-formulation}.
In particular, they do not adequately address:
\begin{itemize}
\item Partial reuse, where only a subset of modules are relevant;
\item Incidental similarity caused by small or generic modules, which should not dominate the overall similarity score.
\end{itemize}

These limitations motivate the design of a new project-wise similarity framework, which we present in the next section.

\section{Proposed Framework}
\label{sec:proposed-framework}

\noindent
To address the limitations discussed in Section~\ref{sec:project-wise-sim}, 
we propose a project-wise similarity framework based on two key ideas:
(1) weighting module-wise similarities to reduce the impact of incidental matches, and
(2) focusing on the most relevant subset of module pairs to capture partial reuse.

\subsection{Symmetric Aggregation}

\noindent
As a preliminary step, we define a simple symmetric aggregation method of module-wise similarities, referred to as \textit{Symmetric Aggregation (SA)}, which serves as the base of our framework.

For a module $x$ and a project $Y$, let
\begin{equation}
\label{eq:top1_noweight}
Top1(x,Y)
=
\max_{y \in Y} sim(x,y)
\end{equation}
denote the highest similarity between $x$ and all modules in $Y$.

Then, the symmetric aggregation (SA) similarity is defined as:

\begin{equation}
Sim_{SA}(P,Q)
=
\frac{
\sum_{i=1}^{n} Top1(p_i,Q)
+
\sum_{j=1}^{m} Top1(q_j,P)
}
{n+m},
\end{equation}
which can also be represented using baseline method equation~\ref{eq:baseline_method}:
\begin{equation} 
    Sim_{SA}(P,Q) = 
    \frac{
        Sim_p(P,Q) + Sim_p(Q,P)
    }
    {2}.
\end{equation}

This aggregation ensures that project-level similarity is independent of the order of inputs, and corresponds to the case where no weighting is applied and all module pairs are considered.

\subsection{Weight Assignment}
\label{sec:weight-assignment}
\noindent
Not all modules contribute equally to plagiarism detection.
In particular, smaller modules are more likely to exhibit incidental similarity~\cite{Lim2009controlFlowEdges}.
To mitigate this effect, we assign weights to modules based on the size of their birthmarks.

Let $K$ denote the number of k-grams in a module.
A naive approach would assign weights proportional to $K$.
However, in projects containing very large modules, such a linear weighting would cause a small number of modules to dominate the overall similarity score, potentially overshadowing the contributions of the remaining modules.

To mitigate this effect, we apply a logarithmic transformation:
\begin{equation}
    W(B(p)) = \ln(K).
\end{equation}

This transformation compresses the range of weights, reducing the dominance of extremely large modules while still preserving the relative importance of module size.

We note that the choice of the weight assignment function is not the primary focus of this study.
While alternative formulations (e.g., $\sqrt{K}$ or linear scaling) are possible, our goal is to demonstrate the effectiveness of incorporating size-based weighting, rather than to optimize the specific functional form.

To normalize weights within a pair of projects $P$ and $Q$, we define:
\begin{equation}
    W_{norm}(p, P, Q) =
    \frac{W(B(p))}{
        \max\bigl( \{W(B(p_i))\} \cup \{W(B(q_j))\} \bigr)
    }.
\end{equation}

Finally, the weighted similarity between two modules is defined as:
\begin{equation}
\begin{split}
 &   sim_{weighted}(p_i, q_j, P, Q) =\\
 &   sim(p_i, q_j) \times \min\bigl(W_{norm}(p_i, P, Q), W_{norm}(q_j, P, Q)\bigr).
\end{split}
\end{equation}

This formulation reduces the influence of smaller modules, which are more prone to incidental similarity.

\subsection{Partial Similarity}
\label{sec:partial-sim}
\noindent
Even with weighting, aggregating all module pairs may dilute the similarity result, especially when only a subset of modules is reused.

To address this, we propose \textit{partial similarity}, an extension of SA, which focuses only on the most relevant subsets of module pairs.

We define $Top1_{weighted}(x,Y)$ as modification of $Top1(x,Y)$ that employs $sim_{weighted}(x,y,X,Y)$ in place of $sim(x,y)$.

Let $S(X,Y)$ denote sets of the highest weighted similarity results for each module in project $X$:
\begin{equation}
    S(X,Y)=\{Top1_{weighted}(x,Y) | \forall x \in X\},
\end{equation}

We sort similarity results in $S$ in descending order and select the top $\alpha\%$ of the highest values:
\begin{equation}
    S_{\alpha}(X,Y) = \text{Top-}\alpha\%\bigl(S(X,Y)\bigr).
\end{equation}

The partial similarity is then defined as:
\begin{equation}
    Sim_{partial}(P,Q,\alpha) =
    \frac{
        \sum S_{\alpha}(P,Q) + \sum S_{\alpha}(Q,P)
    }{
        |S_{\alpha}(P,Q)| + |S_{\alpha}(Q,P)|
    },
\end{equation}

where $\alpha$ is the parameter that controls the comparison scope.

By focusing on high-similarity module pairs, this method emphasizes reused components while ignoring unrelated parts of the projects.

In this study, we evaluate $\alpha \in \{1,5,10,25,50,75\}$.
For $\alpha = 100$, the method is identical to SA. 

\section{Experiment Design}
\label{sec:experiment}

\subsection{Overview and Objectives}
\label{sec:experiment-overview}
\noindent
The objective of this experiment is to evaluate the effectiveness of the proposed project-wise similarity framework from three perspectives:

\begin{itemize}
\item Whether project-wise similarity can reliably distinguish reused and non-reused project pairs;
\item Whether size-based weighting reduces false positives caused by small or generic modules;
\item Whether partial similarity improves detection of partial code reuse.
\end{itemize}

To achieve this, we design an experiment consisting of the following steps:
\begin{itemize}
    \item Dataset preparation and filtering;
    \item Module-wise similarity computation;
    \item Project-wise similarity calculation using multiple methods;
    \item Evaluation using threshold-based classification and harmonic mean (Hmean).
\end{itemize}

\subsection{Dataset}
\label{sec:dataset}
\noindent 
For this study, we constructed a custom dataset consisting of 35 Java projects across 10 categories (Appendix~\ref{sec:appendix-oss-project-overview} (Table~\ref{tab:projects})).

Projects are grouped into categories to ensure that non-reused pairs share similar functionality, making the credibility evaluation more challenging and realistic. 
The categorization was conducted manually based on project tags and descriptions provided in their GitHub repositories.

Different versions of the same project are treated as reused pairs, simulating realistic scenarios in which reused code is modified, extended, or integrated into new contexts.

\medskip

\noindent
The projects were selected from GitHub according to the following criteria:
\begin{itemize}
    \item \textbf{Primary language (Java).} 
    Java must be the dominant language in the project.
    Projects that combine Java with other languages are accepted, but Java must be listed as the primary language on the repository page.  

    \item \textbf{Minimum project size.} 
    Each project (and each selected version) must contain approximately 50 or more Java class files to ensure sufficient structural complexity.

    \item \textbf{Project popularity.} 
    Only projects with at least 50 stars on GitHub were included, to focus on relatively well-maintained and non-trivial software systems.

    \item \textbf{Version availability.} 
    Each project must provide at least four released versions. 
    These versions are used to evaluate resilience by treating newer versions as modified derivatives of earlier ones.
    The process of version selection is described in Appendix~\ref{sec:appendix-oss-project-overview}.
\end{itemize}

\medskip

\noindent
Compiled \textit{.jar} files were obtained either from the project's official release page or from the Maven repository\footnote{\url{https://repo1.maven.org/}} when releases were not directly available.

\medskip

\noindent
For each project version, software birthmarks were extracted using the \textit{pochi} tool. 
Subsequently, module-wise similarity was computed for all relevant module pairs using the similarity functions which will be described in Section~\ref{sec:module-wise-sim-functions}. 

\medskip

\noindent
Overall, this dataset enables systematic evaluation of both resilience (via intra-project version comparisons) and credibility (via inter-project comparisons within the same category), providing a realistic and controlled environment for assessing project-wise similarity methods.

\noindent
To facilitate reproducibility and future research, the dataset used in this study was made publicly available\footnote{The dataset is released at: \url{https://gitlab.com/ou-salab-fedorov/project-wise_comparison_of_software_birthmarks}.}.

\subsection{Dataset Filtering Methods} 
\label{sec:dataset-filtering-methods}
\noindent 
Using \textit{.jar} files as-is is inappropriate for similarity analysis, as they often contain external libraries and small generic modules that do not contribute meaningful information for plagiarism detection. 
Such modules may introduce noise and inflate similarity scores between unrelated projects.

To address these issues, we apply two filtering strategies.

\medskip

\noindent
\textbf{External module filtering}.
While some projects provide dependency-free \textit{.jar} files, most include external libraries. 
We categorize modules into \textit{internal} (compiled from the project's own source code) and \textit{external} (third-party libraries and unrelated files). 
Since external modules are not indicative of code reuse between the inspected projects, including them would introduce irrelevant comparisons (e.g., internal–external or external–external).

Modules are classified using build configuration files such as \textit{pom.xml}\footnote{\url{https://maven.apache.org/pom.html}} (Maven), \textit{build.xml}\footnote{\url{https://ant.apache.org/manual/using.html}} (Ant), and \textit{build.gradle}\footnote{\url{https://docs.gradle.org/current/userguide/build_file_basics.html}} (Gradle). 
Based on this information, all external modules are removed prior to similarity computation.

\medskip

\noindent
\textbf{Logical line count (LLC)-based filtering}.
To reduce false positives caused by small and generic modules, we adopt a modified version of the filtering method used in a past study~\cite{kakimoto2006using}. 
In the original approach, modules with fewer than 30 lines of source code are removed. 
However, raw line counts may include empty lines, comments, and brackets, which do not reflect the actual complexity of the code.

To address this limitation, we employ \textit{logical line count} (LLC), which excludes such non-functional lines. 
Modules with LLC $\leq 30$ are removed from the dataset.

\medskip

\noindent
It is worth noting that this filtering step may appear redundant given the use of size-based weighting in our proposed method. 
However, LLC-based filtering remains necessary for two reasons.

First, removing small modules is not an uncommon practice in software birthmark research~\cite{kakimoto2006using}. 
Such modules are typically generic and contribute little to distinguishing program characteristics, while disproportionately increasing the number of incidental matches. 
Therefore, filtering them out improves the signal-to-noise ratio and reduces computational cost without sacrificing meaningful information.

Second, our experimental evaluation aims to provide a fair comparison between the proposed weighting method and conventional similarity methods. 
If LLC-based filtering were omitted, the conventional methods would be disproportionately affected by small-module noise because they do not explicitly address the influence of small and generic modules. 
As a result, the comparison could unfairly favor the proposed method simply due to differences in noise handling. 
Applying the same filtering condition across all methods therefore provides a more balanced and reliable evaluation framework.

\medskip

\noindent
In Java, nested classes may be compiled into separate class files depending on compiler settings~\cite{ibm2025effects,oracle2025nested}. 
To keep the filtering process simple, nested class files are removed only if their corresponding parent class is also filtered out.

\medskip

\noindent
The effect of the filtering process is summarized in Table~\ref{tab:dataset-filtering}. 
Overall, approximately 82.5\% of modules are removed.

\begin{table*}
\centering
\caption{External module filtering results.}
\begin{tabular}{c c c c}
\toprule
& 
\shortstack{Total \\ (before \\ filtering)} &
\shortstack{After external \\ module filtering} &
\shortstack{After LLC (30) \\ filtering} \\
\midrule
Class file count & 198906 & 57780 & 34855 \\
\bottomrule
\label{tab:dataset-filtering}
\end{tabular}
\end{table*}

\medskip

\noindent
To further validate the filtering criterion, we analyze the relationship between LLC and bytecode size. 
Table~\ref{tab:instructions-per-llc} shows that one logical line corresponds to approximately 2--3 bytecode instructions on average, which were extracted using javap~\footnote{\url{https://docs.oracle.com/javase/9/tools/javap.htm}}.
This suggests that LLC serves as a reasonable proxy for code size.

\begin{table*}
    \caption{Average bytecode instruction counts per single LLC.}
    \label{tab:instructions-per-llc}
    \centering
    \begin{tabular}{c|c}
    \toprule
        \textbf{Subset} & \textbf{Average instruction count per 1 LLC} \\
    \midrule
        Total & 2.707 \\
        LLC $\leq$ 30 & 2.042 \\
        LLC $>$ 30 & 3.33 \\
    \bottomrule
    \end{tabular}
\end{table*}

\medskip

\noindent
Although a large portion of modules is removed, this filtering process improves the overall reliability of similarity computation by eliminating irrelevant and misleading comparisons.
In cases where source code is unavailable, filtering based on approximate instruction counts (e.g., corresponding to LLC thresholds) could be used as an alternative in future work.

\subsection{Compared Methods}
\label{sec:compared-methods}
\noindent 
To evaluate the effectiveness of the proposed approach, we compare it with several baseline and variant methods. 
Each method is designed to highlight different aspects of project-wise similarity computation.

\begin{itemize}

\item \textbf{Baseline method}~\cite{lee2016api}

The method proposed by Lee et al.~\cite{lee2016api}, which computes project similarity based on the average of maximum module-wise similarities.
This method is inherently asymmetric, i.e., $Sim_p(P,Q) \neq Sim_p(Q,P)$.
To mitigate this issue in our experiments, we randomly select either $Sim_p(P,Q)$ or $Sim_p(Q,P)$ and repeat the process 10 times, reporting the average result.

\medskip

\item \textbf{Aggregated similarity}~\cite{fedorov2024comparison}

The method proposed in our previous work~\cite{fedorov2024comparison}, which aggregates the top-$N$ module-wise similarities from both projects.
In this study, we evaluate two configurations: $N=1$ and $N=2$, to examine the effect of considering multiple candidate matches.

\medskip

\item \textbf{Symmetric aggregation (SA)}

The base aggregation method used in the proposed framework, defined as the average of the highest module-wise similarity results between two projects, which also equates to the the average of $Sim_p(P,Q)$ and $Sim_p(Q,P)$.
This method addresses the asymmetry issue of the baseline method without introducing weighting or partial selection.
It corresponds to the unweighted, full-scope version of our proposed framework and serves as a reference point for evaluating the contributions of weights and partial similarity.

\medskip

\item \textbf{Weighted partial similarity}

The proposed method, which integrates both size-based weights and partial similarity.
First, module-wise similarities are adjusted using weights derived from module size.
Then, only the top percentage of similarity values (denoted as \textit{comparison scope}) is selected from both projects, and their overall average is used as the project-wise similarity.

In this study, we evaluate $\text{scope} \in \{1, 5, 10, 25, 50, 75\}$ to analyze the effect of focusing on different subsets of high-similarity module pairs.
% proportions

\medskip

\item \textbf{Random similarity}

A baseline method in which project similarity scores are assigned randomly over the course of 100 iterations and then rounded to either 0 or 1.
This serves as a lower bound for performance, representing the expected outcome without any meaningful similarity computation.

\end{itemize}

These methods collectively enable a systematic evaluation of the contributions of symmetry, weighting, and partial selection in project-wise similarity computation.

\subsection{Module-wise Similarity Functions}
\label{sec:module-wise-sim-functions}

\noindent
Project-wise similarity is computed by aggregating similarities between matched module pairs.
Therefore, the choice of module-wise similarity function can directly affect the final results.

In this study, we consider the following representative similarity functions:

\begin{itemize}
\item Cosine similarity (count vectorization);
\item Cosine similarity (TF-IDF vectorization);
\item Dice index;
\item Jaccard similarity coefficient;
\item Simpson similarity index;
\item Edit distance.
\end{itemize}

For cosine similarity, birthmarks are vectorized using two commonly used approaches (count and TF-IDF vectorization) implemented in \textit{scikit-learn}: CountVectorizer\footnote{\url{https://scikit-learn.org/stable/modules/generated/sklearn.feature_extraction.text.CountVectorizer.html}} and TfidfVectorizer\footnote{\url{https://scikit-learn.org/stable/modules/generated/sklearn.feature_extraction.text.TfidfVectorizer.html}}.

These similarity functions are treated as experimental factors, and their impact on project-wise similarity is analyzed in Section~\ref{sec:res-sim-func-effect}.

\subsection{Evaluation Metrics}
\label{sec:evaluation-metrics}
\noindent
To evaluate the effectiveness of project-wise similarity measures, we extend the concepts of \textit{resilience} and \textit{credibility}, originally defined for module-level birthmarks, to the project level.

\paragraph{Project-wise similarity result evaluation}
Given a project-wise similarity score $Sim(P,Q)$, we determine whether two projects are considered similar using a threshold $\mathcal{E}$:
\begin{equation}
    Sim(P,Q)
    =
    \begin{cases}
        \leq \mathcal{E} & \text{not similar},\\
        > \mathcal{E} & \text{similar}.
    \end{cases}
\end{equation}

\paragraph{Evaluation sets}
To evaluate different aspects of performance, we define two sets of project pairs:

\begin{itemize}
\item \textbf{Reused pairs} ($P_{\text{reused}}$):  
Pairs of different versions of the same project. 
These simulate realistic reuse scenarios where code is modified and extended.
\item \textbf{Non-reused pairs} ($P_{\text{not reused}}$):  
Pairs of different projects within the same category. 
These pairs share similar functionality but are independently developed, making the task of distinguishing them more challenging.
\end{itemize}

Each set is further decomposed as:
\begin{equation}
P_{\text{reused}} = TP + FN, \quad
P_{\text{not reused}} = TN + FP,
\end{equation}
where $TP$, $FN$, $TN$, and $FP$ denote true positives, false negatives, true negatives, and false positives, respectively.

\paragraph{Resilience and credibility}
We define two evaluation metrics:

\begin{equation}
\text{resilience rate}
=
\frac{TP}{TP + FN},
\end{equation}

\begin{equation}
\text{credibility rate}
=
\frac{TN}{TN + FP}.
\end{equation}

The resilience rate measures the ability to correctly identify reused projects despite modifications, while the credibility rate measures the ability to avoid false positives among independently developed projects.

\paragraph{Threshold selection}
The performance of a similarity measure depends on the choice of threshold $\mathcal{E}$.
In this study, rather than fixing a predefined threshold, we determine the optimal threshold for each project category.

Specifically, for each category, we perform an exhaustive search over $\mathcal{E} \in (0.0, 1.0)$ with a step size of $0.001$, and select the value that maximizes the harmonic mean of resilience and credibility:

\begin{equation}
\text{Hmean}_{\text{category}}
=
\frac{2}{
\frac{1}{\text{resilience rate}}
+
\frac{1}{\text{credibility rate}}
}.
\end{equation}

This procedure is intended to evaluate the \textit{relative effectiveness} of similarity measures under their best achievable conditions, rather than to simulate a deployment scenario where the threshold must be fixed in advance.
Due to the limited number of projects per category, splitting the dataset into separate training and evaluation subsets is not feasible without significantly reducing statistical reliability.
Therefore, the reported results should be interpreted as an upper-bound comparison of the methods, focusing on their intrinsic capability to balance resilience and credibility.

\paragraph{Overall evaluation}
Using the optimal threshold for each category, we compute the final performance across all categories using the macro-level harmonic mean:

\begin{equation}
\text{Hmean}
=
\frac{2k}{
\sum_{i=1}^{k}
\left(
\frac{1}{\text{resilience rate}_i}
+
\frac{1}{\text{credibility rate}_i}
\right)
},
\end{equation}
where $k$ is the number of categories.

\paragraph{Rationale}
We adopt the harmonic mean because it penalizes imbalanced performance: a method that achieves high resilience but low credibility (or vice versa) will receive a low score.  
This is particularly important in software reuse detection, where both avoiding missed detections (FN) and preventing false alarms (FP) are critical.

Compared to metrics such as F1-score, which focuses on a single classification setting, the proposed Hmean explicitly balances the two complementary objectives of reuse detection: robustness to modification and discrimination between independent projects.

\section{Experimental Results}
\label{sec:results}

\subsection{Overall Comparison of Methods}
\label{sec:res-overall-comparison}
\begin{figure*}
    \centering
    \includegraphics[width=0.7\linewidth]{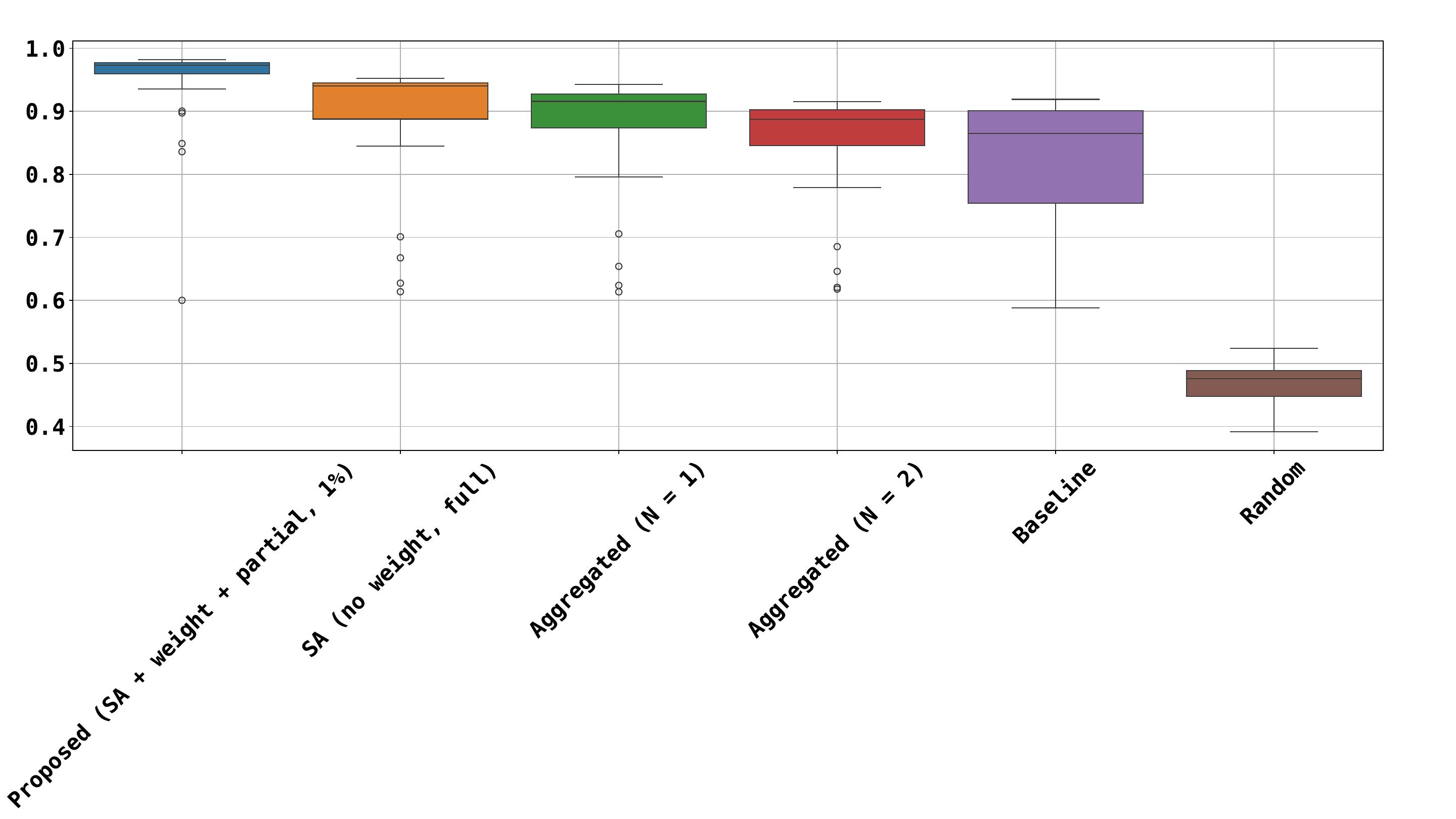}
    \caption{Overall comparison of project-wise similarity methods in terms of Hmean. For the proposed method, the best scope (1\%) is used, while results are aggregated over all $k$-gram sizes and module-wise similarity functions to ensure a fair comparison.}
    \label{fig:overall}
\end{figure*}

\noindent
Fig.~\ref{fig:overall} presents the overall comparison of project-wise similarity methods in terms of Hmean.
The results are aggregated over all configurations of $k$-gram sizes and module-wise similarity functions to ensure a fair comparison without favoring any specific parameter setting.
For the method incorporating partial similarity, the best-performing scope (identified in Section~\ref{sec:res-partial-sim-effect}) is used.

The results show a clear ranking of methods.
The proposed method (weighting + partial similarity, scope = 1\%) achieves the highest performance, with a noticeable margin over all other methods.
In addition to its superior median Hmean, it also exhibits relatively low variance, indicating stable performance across different configurations (Table~\ref{tab:overall-hmean-stats}).

The symmetric aggregation method without weighting or partial similarity (SA (no weight)) performs consistently better than the existing approaches, including aggregated similarity and the baseline method.
This result suggests that the symmetric formulation itself provides a strong foundation for project-wise similarity, even without additional enhancements.
Although SA (no weight) is conceptually similar to aggregated similarity with $N=1$, it is simpler in formulation while achieving comparable or better performance, making it a suitable baseline for further extensions.

Comparing the aggregated similarity variants, $N=1$ outperforms $N=2$.
This indicates that introducing the parameter $N$ to consider cases where the original module is plagiarized by multiple modules and to lower the impact of false positives, does not necessarily improve overall performance in this setting.
A possible explanation is that higher values of $N$ might also increase false negatives when trying to reduce the false positives, resulting in lower overall Hmean.
Since this method is not the focus of this study, we do not investigate this effect further.

The baseline method shows lower performance and larger variance compared to both SA and aggregated similarity.
This can be attributed to its asymmetric nature, which introduces instability even when averaging over multiple runs.

As expected, the random similarity method performs significantly worse than all other methods, confirming that the evaluation framework is meaningful.

We conduct Wilcoxon signed-rank test using the results from Proposed (weight + partial, 1\%) and Aggregated (N = 1) methods.
$p = 3.0496 \times 10^{-5} < 0.01$, thus, the difference in performance is statistically significant.

Overall, these results demonstrate that the proposed method clearly outperforms existing approaches, and that the symmetric aggregation formulation provides a strong and effective basis for project-wise similarity.

\begin{table*}
\centering
\caption{Average, median Hmean and variance per project-wise similarity method.}
\label{tab:overall-hmean-stats}
\begin{tabular}{c | c c c}
\toprule
\textbf{Method} &
\textbf{Avg. Hmean} & 
\textbf{Median Hmean} &
\textbf{Variance} \\
\midrule
Proposed (weight + partial, 1\%) & 0.9497 $\pm$ 0.0695 & 0.9728 & 0.0048 \\
SA (no weight) & 0.8968 $\pm$ 0.0929 & 0.94 & 0.0086 \\
Aggregated (N = 1) & 0.8779 $\pm$ 0.0896 & 0.9157 & 0.008 \\
Aggregated (N = 2) & 0.8527 $\pm$ 0.0837 & 0.8875 & 0.007 \\
Baseline & 0.8145 $\pm$ 0.1105 & 0.8645 & 0.0122 \\
Random & 0.4707 $\pm$ 0.0288 & 0.4757 & 0.0008 \\
\bottomrule 
\end{tabular}
\end{table*}

\subsection{Effect of Weighting}
\label{sec:res-weighting-effect}
\noindent
Fig.~\ref{fig:weight} shows the effect of weighting on the proposed method (partial similarity, scope = 1\%) in terms of Hmean.

\begin{singlecolelement}
    \centering
    \includegraphics[width=\linewidth]{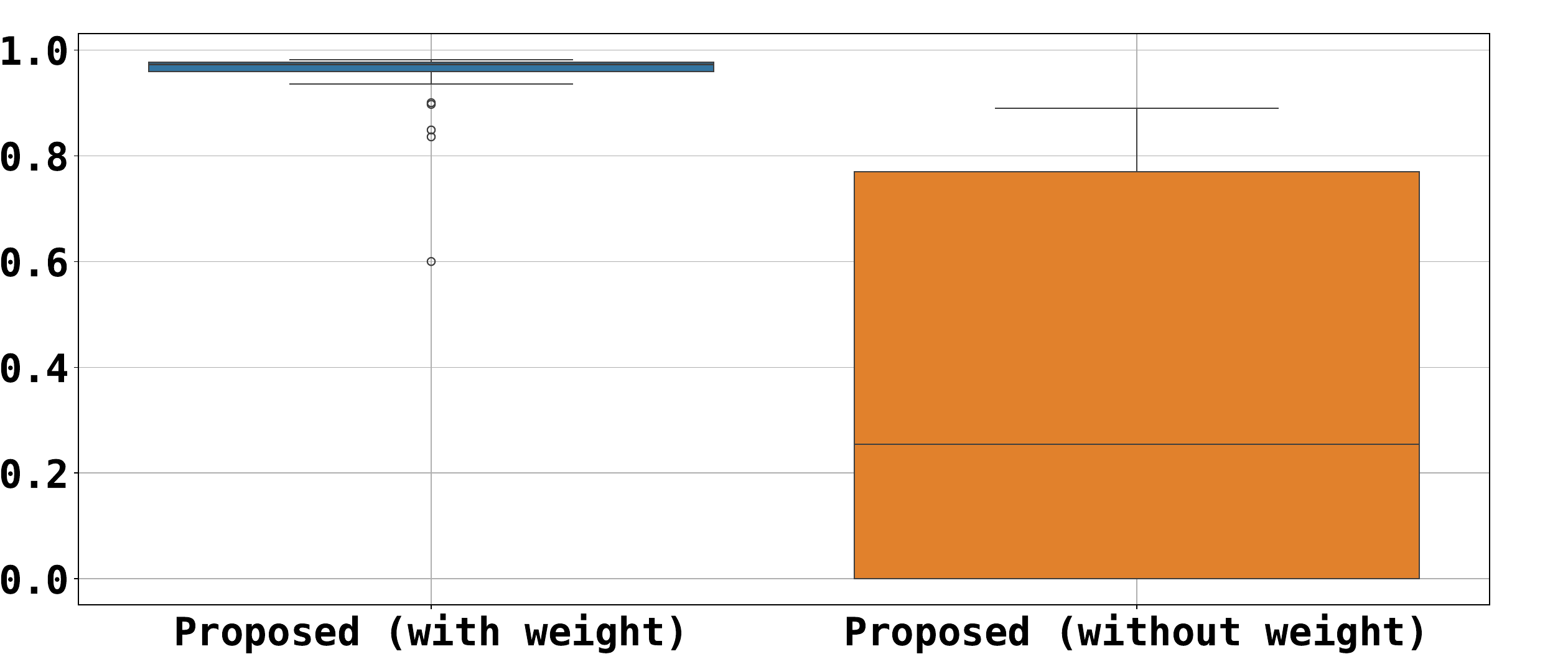}
    \captionof{figure}{Effect of weighting on the proposed method (partial similarity, scope = 1\%) in terms of Hmean.}
    \label{fig:weight}
\end{singlecolelement}

The results demonstrate that weights has a substantial impact on performance.
When weights are applied, the method achieves significantly higher Hmean values with relatively low variance, indicating both strong accuracy and stability across configurations.
In contrast, without weighting, the performance drops dramatically and exhibits very large variability.
This improvement is statistically significant based on the Wilcoxon signed-rank test ($p = 1.6766 \times 10^{-7} < 0.01$).

Notably, the median performance without weight assignment is even lower than that of the random similarity method shown in Fig.~\ref{fig:overall}.
This result is particularly striking, given that the SA without weighting (Fig.~\ref{fig:overall}) performed as the second-best method.
This indicates that combining partial similarity (scope = 1\%) with an unweighted formulation leads to a severe degradation in performance.

A plausible explanation is that, under partial similarity, only the top fraction of module pairs is considered.
Without weights, these top-ranked pairs are often dominated by small modules, which tend to exhibit incidental similarity even when no actual code reuse exists.
As a result, false positives increase significantly, leading to unreliable project-wise similarity scores.

In contrast, the weighting scheme suppresses the influence of small modules by assigning them lower importance.
This prevents such modules from dominating the top-ranked pairs, thereby reducing false positives and stabilizing the results.

These findings suggest that partial similarity alone is insufficient and may even be harmful without an appropriate mechanism for assignment of weights.
Instead, weighting and partial similarity should be considered as complementary components that jointly enable robust project-wise similarity estimation.

\begin{singlecolelement}
    \centering
    \includegraphics[width=\linewidth]{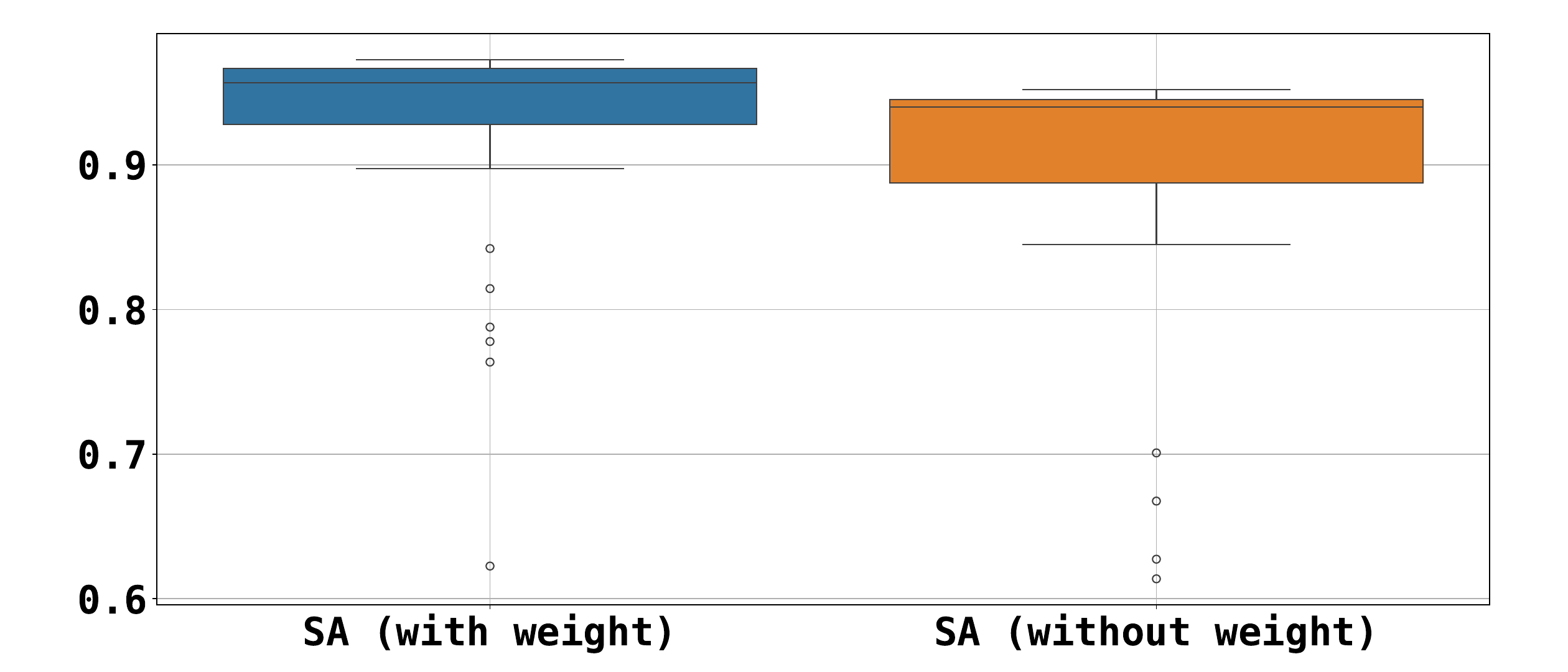}
    \captionof{figure}{Effect of weighting on the SA in terms of Hmean.}
    \label{fig:weight-sa-full}
\end{singlecolelement}

To isolate the effect of weighting from that of partial similarity, we additionally evaluate the proposed method under the condition of \textit{full} scope (i.e., scope = 100\%), which is equal to SA.
The results, shown in Fig.~\ref{fig:weight-sa-full}, indicate that weight assignment consistently improves performance even without partial similarity. 
Specifically, the weighted variant achieves higher Hmean values and exhibits slightly lower variance compared to the non-weighted version.
The difference is also statistically significant according to the Wilcoxon signed-rank test ($p = 0.0121 < 0.05$):

These findings confirm that weighting alone contributes positively to performance by reducing the influence of incidental similarities from small modules. At the same time, when combined with partial similarity, its effect becomes substantially more pronounced, as previously observed.

\subsection{Effect of Partial Similarity}
\label{sec:res-partial-sim-effect}
\begin{figure*}
    \centering
    \includegraphics[width=0.7\linewidth]{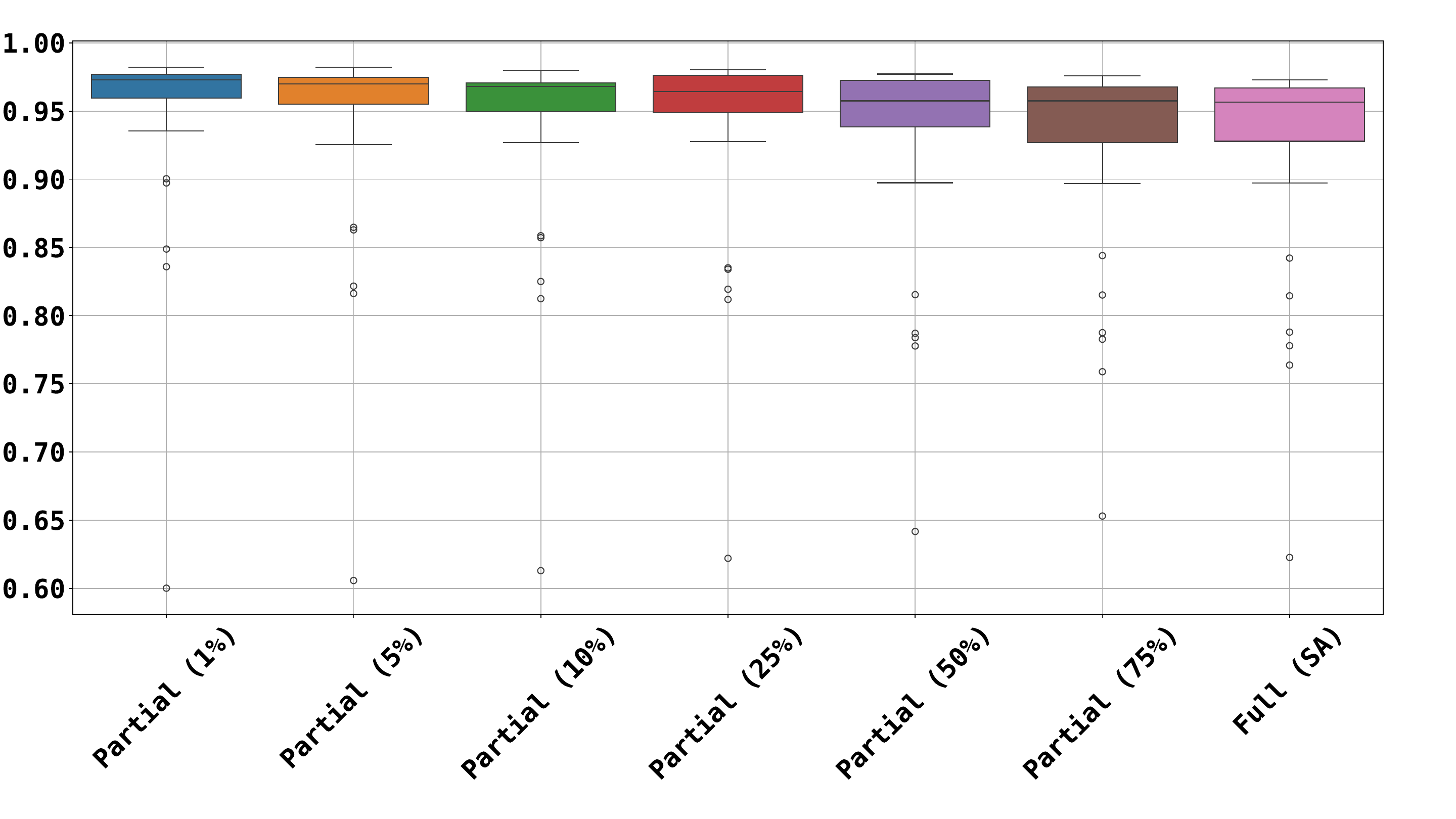}
    \caption{Effect of comparison scope (partial similarity ratio) on the proposed method (partial similarity with weighting), including the full-scope case (100\%) (SA), evaluated using Hmean.}
    \label{fig:partial}
\end{figure*}

\noindent
Fig.~\ref{fig:partial} illustrates the effect of the comparison scope on the performance of partial similarity with weighting.
The comparison scope controls the proportion of top similarity values used in the aggregation.

The results indicate that partial similarity improves performance compared to using all similarity values (i.e., scope = 100\%).
Smaller scope values generally lead to higher Hmean, suggesting that focusing on the most similar module pairs helps emphasize meaningful reuse signals.

Small scope values (e.g., 1\% and 5\%) consistently achieve the highest performance, with 1\% often yielding the best results.
While larger scope values (10--75\%) produce similar performance levels, their improvements over the full-scope setting are less pronounced.

We also conduct a Wilcoxon signed-rank test comparing the best-performing scope value (i.e., scope = 1\%) with the full-scope setting (scope = 100\%).
The results show statistically significant improvements ($p = 0.0095 < 0.01$), supporting the effectiveness of partial similarity.

These results suggest that partial similarity serves as an effective complementary mechanism to weighting.
Also, focusing on a very small subset of highly similar module pairs is effective for capturing reuse signals.

\subsection{Effect of k-gram Size}
\label{sec:res-k-gram-effect}
\begin{figure*}
    \centering
    \includegraphics[width=0.7\linewidth]{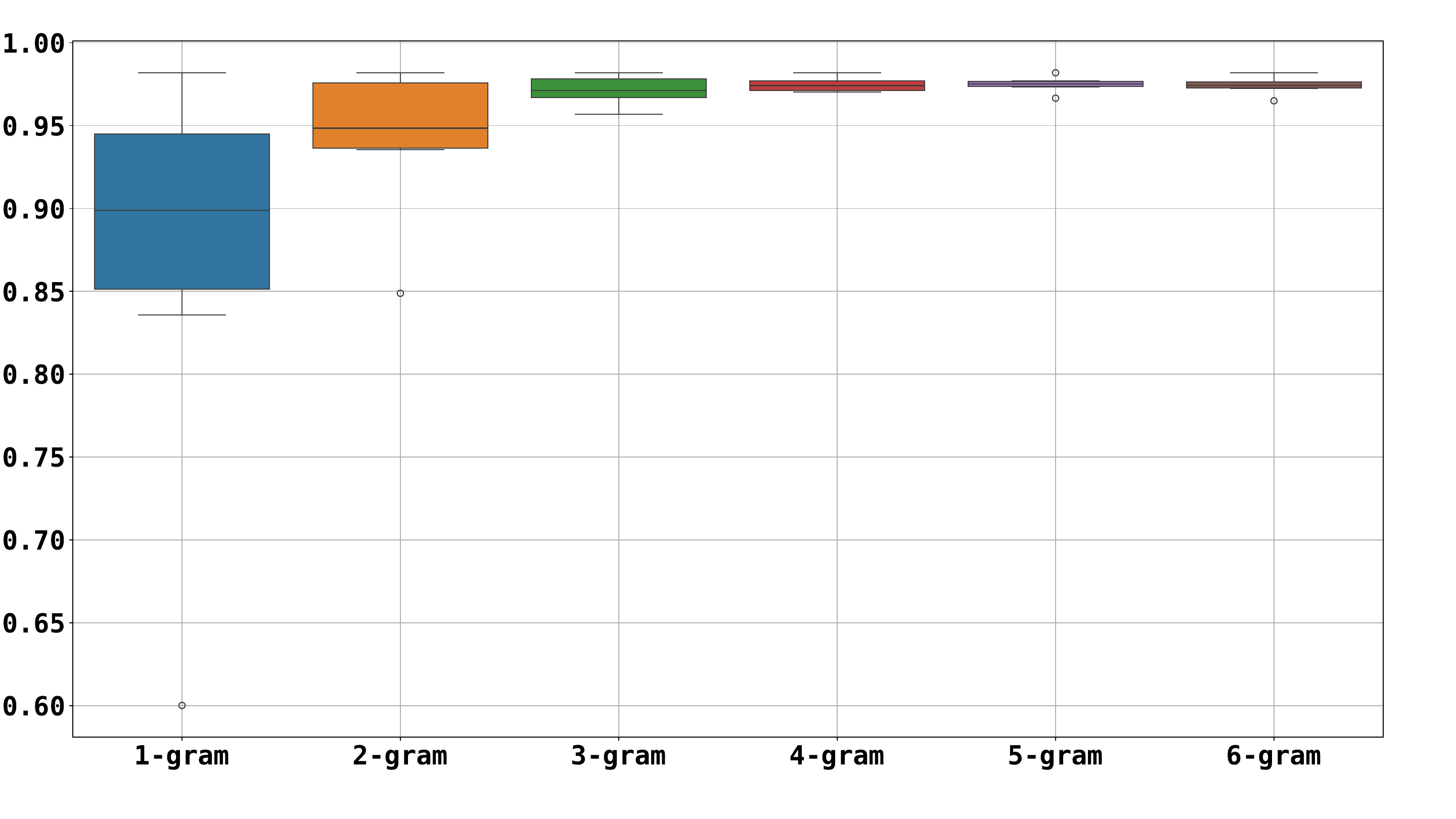}
    \caption{Effect of $k$-gram size ($k \in \{1,\ldots,6$\}) on the proposed method (weighting + partial similarity, scope = 1\%), evaluated in terms of Hmean.}
    \label{fig:kgram}
\end{figure*}

\noindent
Fig.~\ref{fig:kgram} shows the impact of different $k$-gram sizes on the performance of the proposed method.

The results indicate that performance improves as $k$ increases from 1 to 4, after which the improvement saturates.
This behavior suggests that smaller values of $k$ do not capture sufficient structural information, while larger values provide diminishing returns.

Based on these observations, values $k \in \{3, 4\}$ appear to be a practical choice for achieving a good balance between performance and computational cost.

\subsection{Effect of Similarity Function}
\label{sec:res-sim-func-effect}
\begin{figure*}
    \centering
    \includegraphics[width=0.7\linewidth]{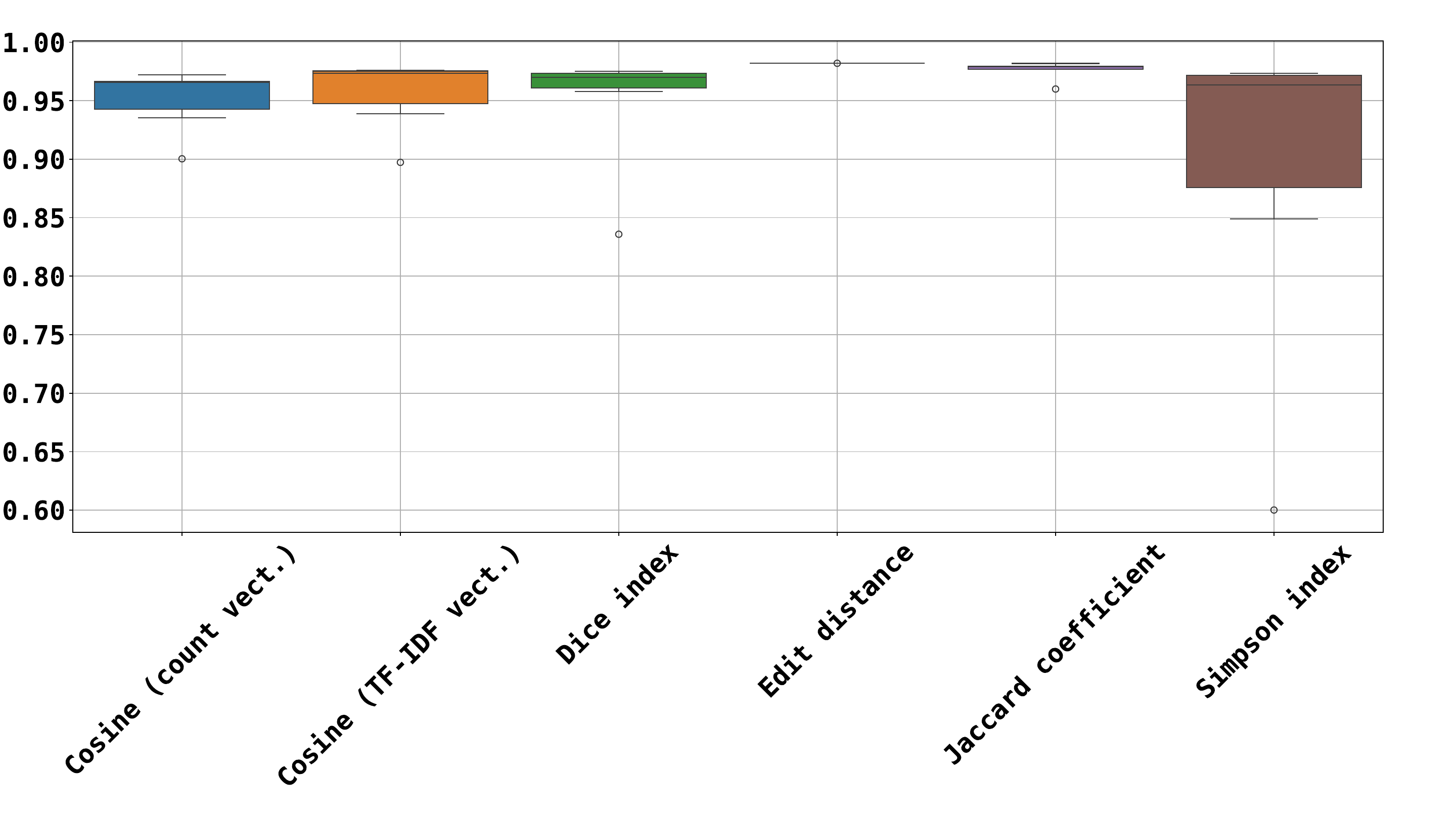}
    \caption{Comparison of module-wise similarity functions for proposed method (weighting + partial similarity, scope = 1\%), evaluated in terms of Hmean.}
    \label{fig:simfunc}
\end{figure*}

\noindent
Fig.~\ref{fig:simfunc} presents the comparison of module-wise similarity functions for the proposed method (weighting + partial similarity, scope = 1\%), evaluated in terms of Hmean.

The results clearly show that edit distance achieves the highest performance among all evaluated similarity functions.
In addition to its superior median Hmean, edit distance also exhibits remarkably low variance, indicating stable performance across different configurations.

The Jaccard coefficient provides the second-best performance, with relatively high Hmean values and moderate variability.
In contrast, the remaining similarity functions show both lower performance and substantially higher variance, suggesting sensitivity to parameter settings and reduced robustness.

In particular, the Simpson index performs the worst among all evaluated functions.
It not only yields lower Hmean values but also demonstrates very large variability, indicating unstable and unreliable behavior in this context.

Overall, these results suggest that edit distance is the most suitable similarity function for module-wise comparison in the proposed framework, providing both high accuracy and strong robustness.

\subsection{Ranking-Based Analysis Across Configurations}
\label{sec:res-ranking}
\noindent
Table~\ref{tab:ranking} presents the top-ranked configurations based on Hmean across all parameter combinations.

\begin{table*}
\centering
\ssmall
\caption{Top-30 configurations ranked by Hmean across all combinations of methods and parameter settings, including $k$-gram size, comparison scope, and module-wise similarity function.}
\label{tab:ranking}
\begin{tabular}{c | c c c c}
\toprule
\textbf{Rank} &
\textbf{Method} &
\textbf{Birthmark} & 
\textbf{Sim. function} &
\textbf{Hmean} \\
\midrule
1 & Partial (5\%) & 5-gram & Edit distance & 0.9823 \\
2 & Partial (1\%) & 1-gram & Edit distance & 0.982 \\
3 & Partial (5\%) & 4-gram & Edit distance & 0.982 \\
3 & Partial (5\%) & 6-gram & Edit distance & 0.982 \\
3 & Partial (5\%) & 3-gram & Edit distance & 0.982 \\
4 & Partial (1\%) & 5-gram & Edit distance & 0.9819 \\
4 & Partial (1\%) & 2-gram & Edit distance & 0.9819 \\
4 & Partial (1\%) & 3-gram & Edit distance & 0.9819 \\
4 & Partial (1\%) & 6-gram & Edit distance & 0.9819 \\
4 & Partial (1\%) & 4-gram & Edit distance & 0.9819 \\
5 & Partial (1\%) & 2-gram & Jaccard coefficient & 0.9819 \\
6 & Partial (5\%) & 1-gram & Edit distance & 0.9816 \\
7 & Partial (25\%) & 3-gram & Edit distance & 0.9802 \\
8 & Partial (10\%) & 1-gram & Edit distance & 0.9801 \\
9 & Partial (1\%) & 3-gram & Jaccard coefficient & 0.9799 \\
10 & Partial (5\%) & 2-gram & Edit distance & 0.9799 \\
11 & Partial (25\%) & 5-gram & Edit distance & 0.9794 \\
12 & Partial (25\%) & 1-gram & Edit distance & 0.9793 \\
13 & Partial (25\%) & 3-gram & Jaccard coefficient & 0.9787 \\
13 & Partial (25\%) & 2-gram & Edit distance & 0.9787 \\
14 & Partial (10\%) & 2-gram & Jaccard coefficient & 0.9786 \\
15 & Partial (10\%) & 2-gram & Edit distance & 0.9784 \\
16 & Partial (25\%) & 4-gram & Edit distance & 0.9773 \\
16 & Partial (25\%) & 4-gram & Jaccard coefficient & 0.9773 \\
17 & Partial (1\%) & 4-gram & Jaccard coefficient & 0.9773 \\
18 & Partial (50\%) & 3-gram & Edit distance & 0.9771 \\
19 & Partial (10\%) & 3-gram & Edit distance & 0.977 \\
20 & Partial (5\%) & 3-gram & Jaccard coefficient & 0.9769 \\
20 & Partial (1\%) & 6-gram & Jaccard coefficient & 0.9769 \\
20 & Partial (1\%) & 5-gram & Jaccard coefficient & 0.9769 \\
21 & Partial (5\%) & 4-gram & Jaccard coefficient & 0.9769 \\
22 & Partial (5\%) & 2-gram & Jaccard coefficient & 0.9766 \\
22 & Partial (25\%) & 6-gram & Jaccard coefficient & 0.9766 \\
23 & Partial (25\%) & 6-gram & Edit distance & 0.9765 \\
24 & Partial (50\%) & 5-gram & Edit distance & 0.9765 \\
25 & Partial (50\%) & 2-gram & Edit distance & 0.9761 \\
26 & Partial (1\%) & 5-gram & Cosine (TF-IDF vect.) & 0.976 \\
27 & Partial (1\%) & 4-gram & Cosine (TF-IDF vect.) & 0.976 \\
28 & Partial (50\%) & 4-gram & Jaccard coefficient & 0.976 \\
29 & Partial (25\%) & 6-gram & Dice index & 0.9759 \\
30 & Partial (50\%) & 4-gram & Edit distance & 0.9758 \\
30 & Partial (75\%) & 4-gram & Jaccard coefficient & 0.9758 \\
\bottomrule 
\end{tabular}
\end{table*}

The ranking results reveal that the highest-performing configurations are consistently dominated by the proposed method (SA + weighting + partial similarity).
In particular, configurations with small comparison scopes (e.g., 1\% and 5\%) frequently appear at the top of the ranking.
Additionally, edit distance is the most commonly used similarity function among the top-ranked configurations.

Notably, non-SA methods do not appear in the top 30 rankings, indicating that the superiority of the proposed framework is not limited to specific parameter choices.

These results demonstrate that the proposed method is robust across a wide range of parameter configurations.

\subsection{Per-category Analysis}
\label{sec:res-per-category-analysis}
\begin{figure*}
    \centering
    \includegraphics[width=0.7\linewidth]{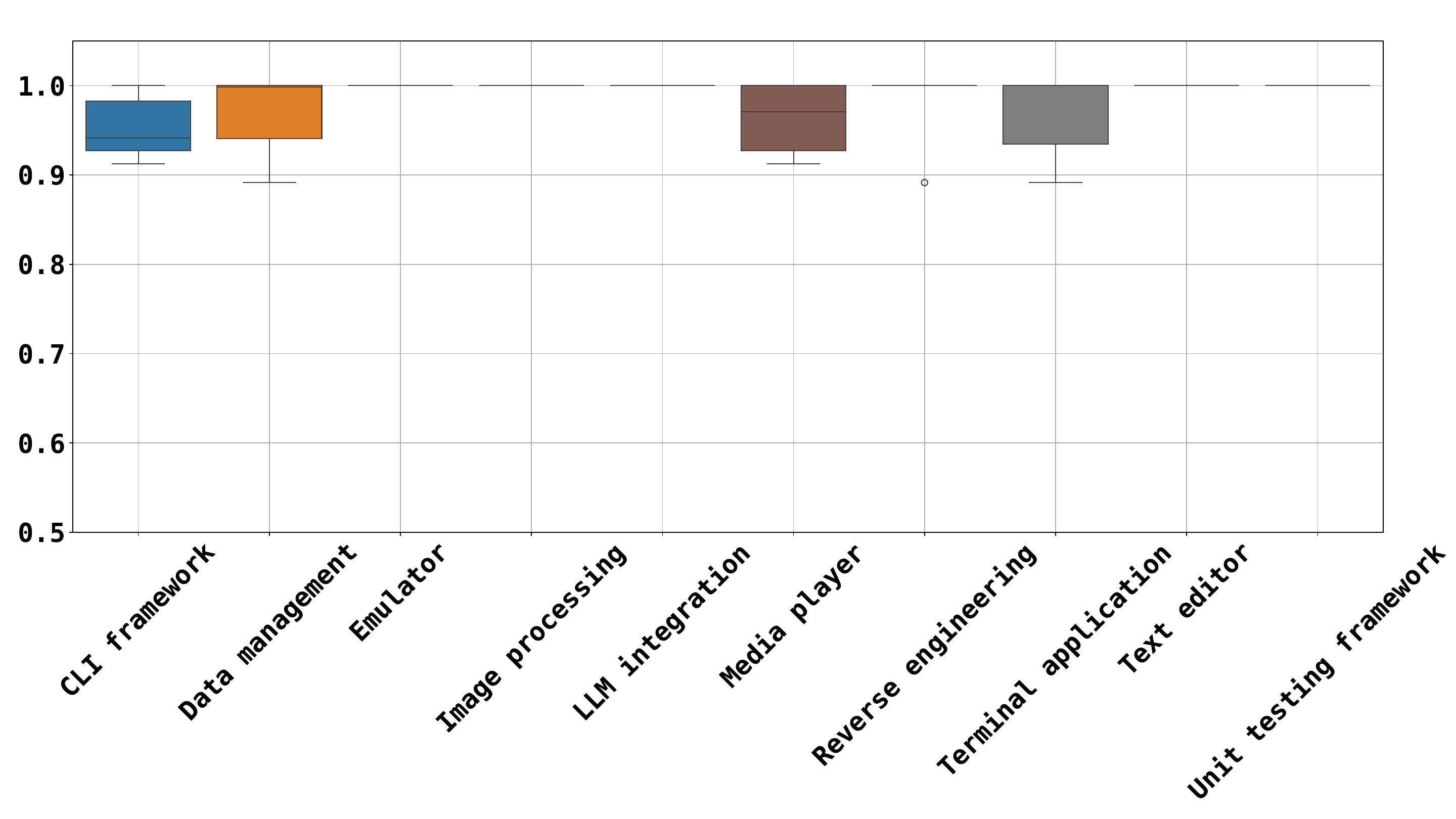}
\caption{Per-category performance of the proposed method (weighting + partial similarity) in terms of Hmean.
Results are shown for configurations with $k \in \{4,5,6\}$ and $\text{scope} \in \{1\%, 5\%\}$ using edit distance similarity function.}
    \label{fig:category}
\end{figure*}

\noindent
Fig.~\ref{fig:category} presents the performance of the proposed method across different project categories.
The analysis focuses on configurations that achieved strong performance in previous sections, namely $k \in \{4,5,6\}$ and $\text{scope} \in \{1\%, 5\%\}$, using edit distance as the module-wise similarity function.

The results show that the proposed method consistently achieves high Hmean values across all categories, indicating strong overall performance.
While some variation can be observed depending on the category, the differences are relatively moderate and no category exhibits a significant degradation in performance.

These observations suggest that the proposed method is robust to differences in application domains and does not rely on category-specific characteristics.
In particular, the consistently high performance across diverse categories supports the general applicability of the method for project-wise similarity estimation.

Overall, the results confirm that the proposed framework maintains both effectiveness and stability across heterogeneous software categories.

\subsection{Summary of Findings}
\label{sec:results-summary}
\noindent
The experimental results can be summarized as follows.

First, the proposed symmetric aggregation (SA) achieves better performance compared to the existing methods, even without additional enhancements (weight assignment and partial similarity), indicating that the symmetric formulation itself provides a strong baseline.

Second, weight assignment plays a crucial role in improving performance by reducing the impact of incidental similarities, particularly those arising from small modules. 
While weighting alone already leads to measurable improvements, its effect becomes substantially more pronounced when combined with partial similarity.

Third, partial similarity further enhances the results by focusing on the most relevant subset of module pairs. 
However, its effectiveness strongly depends on the presence of weighting; without it, partial selection may amplify noise and degrade performance. 
This indicates that weight assignment and partial similarity are complementary mechanisms that jointly contribute to robust similarity estimation.

Finally, the proposed method demonstrates robustness across different parameter settings and project categories, consistently achieving high performance with low variance.

These findings confirm the effectiveness and practical applicability of the proposed project-wise similarity framework.

\section{Threats to Validity}
\label{sec:validity-threats}
\noindent
This section discusses potential limitations of the study and their possible impact on the interpretation and generalization of the results.

\subsection{Dataset-related Threats}

\begin{itemize}

\item \textbf{Assumption of reuse and non-reuse pairs}: 
In this study, different versions of the same project are treated as reused pairs, while different projects within the same category are treated as non-reused pairs.
This design provides a practical and reproducible approximation of project-level software reuse scenarios, enabling systematic evaluation across many configurations.

However, the dataset does not contain confirmed real-world plagiarism cases or adversarially modified software.
Therefore, the evaluated reuse patterns may not fully represent all forms of unauthorized code reuse observed in practice.
Constructing and evaluating datasets based on real-world plagiarism cases remains an important direction for future work.

\item \textbf{Dataset construction and selection}: 
The dataset consists of Java open-source projects selected from GitHub repositories under several filtering conditions, including project size, availability of executable artifacts, and category alignment.
Although these conditions were necessary to enable systematic comparison, they may introduce biases that do not fully reflect the diversity of real-world software ecosystems.

In addition, the dataset mainly contains established projects with over 4 or more official releases across the lifetime and a minimum of 50 stars on GitHub.
The observed similarity characteristics may therefore differ from those of smaller, proprietary, or less actively maintained systems.

\item \textbf{Temporal diversity of project versions}: 
The analyzed project versions span a long time period, from 2002 to 2025.
During this period, software development practices, frameworks, and tooling have significantly evolved.
Such temporal differences may affect project structure and similarity characteristics across versions.

\end{itemize}

\subsection{Evaluation-related Threats}

\begin{itemize}

\item \textbf{Threshold-based evaluation}: 
The proposed methods are evaluated using threshold-based binary classification derived from project-wise similarity scores.
The threshold for each project category is selected to maximize the harmonic mean of resilience and credibility.

This evaluation strategy is intended to compare the relative effectiveness of similarity measures under optimized conditions, rather than to estimate directly deployable real-world detection accuracy.
Because the dataset size is limited, the study does not perform separate threshold training and testing procedures.
Therefore, the reported results should be interpreted primarily as comparative performance estimates between methods.

In addition, alternative threshold-independent evaluation strategies, such as ranking-based metrics or ROC/AUC analysis, were not investigated in this study and remain part of future work.

\item \textbf{Parameter sensitivity}: 
The performance of project-wise similarity depends on several parameters, including the $k$-gram size, module filtering conditions, weighting scheme, and partial similarity scope.
Although extensive experiments across many parameter configurations were conducted, different parameter ranges or alternative formulations may lead to different performance characteristics.

\end{itemize}

\subsection{Generalization-related Threats}

\begin{itemize}

\item \textbf{Programming language and birthmark scope}: 
The evaluation focuses exclusively on Java software projects and static $k$-gram birthmarks.
Therefore, the findings may not directly generalize to other programming languages, software ecosystems, or alternative birthmark representations. 

\item \textbf{Scope of reuse scenarios}: 
The study focuses on project-wise similarity under partial reuse scenarios, where reused modules are embedded within larger software systems.
The experiments do not explicitly evaluate robustness against strong obfuscation or adversarial code transformations.
Such scenarios are outside the scope of the current study and should be investigated separately in future work.

\end{itemize}

\noindent
Despite these limitations, the study provides a comprehensive empirical evaluation across diverse datasets, parameter configurations, and similarity formulations, offering useful insights into project-wise similarity measurement for software birthmarks.

\section{Conclusion}
\label{sec:conclusion}
\noindent
In this paper, we addressed the need for \textit{project-wise} similarity in software birthmark comparison, which serves as a retrieval step for identifying candidate projects that may contain reused or plagiarized code prior to detailed analysis.

We proposed a project-wise similarity framework based on symmetric aggregation of module-level similarities, and introduced two complementary mechanisms: size-based weight assignment and partial similarity.
While the symmetric aggregation provides a simple and stable foundation, our focus was on improving robustness against two key challenges in project-level comparison: incidental similarity from small modules and partial code reuse.

Through extensive experiments, we obtained several important findings.
First, symmetric aggregation alone already provides a strong and reliable baseline, performing competitively with or better than existing approaches.
Second, weighting plays a critical role in improving performance by suppressing the influence of small, noise-prone modules.
Third, partial similarity further enhances detection by focusing on the most relevant subset of module pairs; however, its effectiveness strongly depends on the presence of weights.
Without weighting, partial selection may amplify noise and degrade performance, whereas their combination yields substantial improvements.
These results indicate that weight assignment and partial similarity are complementary mechanisms that jointly enable robust project-wise similarity estimation.

In addition, we observed that configurations with small scopes (e.g., 1\%, 5\%), higher $k$-gram sizes, and edit distance similarity function tend to provide stable and high performance across diverse project categories.

Overall, the results demonstrate that, in the inspected context, in the absence of obfuscation (and other transformations) in the compared projects, effective project-wise similarity does not require complex matching strategies.
Instead, a simple combination of symmetric aggregation, size-aware weights, and selective focus on high-similarity module pairs provides a robust and practical solution.

As future work, we plan to extend the evaluation to other programming languages and larger-scale datasets, investigate robustness against obfuscation and adversarial transformations, and explore alternative weighting strategies and similarity representations, as well as expanding to other types of birthmarks.

\appendix

\section{Overview of the Used Open-source Software Projects}
\label{sec:appendix-oss-project-overview}
\noindent 
Table~\ref{tab:projects} lists all projects used in this study, including their categories and selected versions.

For each project, four versions were selected to evaluate resilience under functional evolution: 
the earliest and the latest available versions (as of June 18, 2025), along with two intermediate versions sampled at approximately equal intervals from the version history. 
This design provides a balanced coverage of both short-term and long-term code evolution.

To ensure sufficient structural complexity for meaningful comparison, only versions containing approximately 50 or more Java class files were considered. 
If a selected version did not provide an available \textit{.jar} file (either on GitHub or the Maven repository), the closest alternative version was used instead.

The selection process was automated using a Python script to ensure consistency.

For projects containing multiple \textit{.jar} files in a given version, all available project components were included. 
Note that external dependencies are subsequently removed during the filtering stage described in Section~\ref{sec:dataset-filtering-methods}, ensuring that only project-internal modules are analyzed.

Each version is represented in the format 
\textless version name\textgreater\_\textless release date in ISO 8601~\cite{kuhn-iso-time} format\textgreater.

\begin{table*}
\centering
\ssmall
\caption{Open-source projects.}
\label{tab:projects}
\begin{tabular}{c | c | c c c c}
\toprule
\textbf{Category} & 
\textbf{Project name} &
\textbf{Version 1} &
\textbf{Version 2} &
\textbf{Version 3} &
\textbf{Version 4} \\
\midrule

\multirow{4}{*}{\shortstack{CLI\\ framework}} 
        &
        Airline & 0.4\_2012-8-21 & 0.5\_2013-01-10 & 0.7\_2014-11-06 & 0.9\_2019-12-06 \\
        &
        JCommander & 1.30\_2012-10-27 & 1.48\_2015-04-10 & 1.69\_2017-04-17 & 1.82\_2022-01-10 \\
        &
        picocli & 0.9.0\_2017-04-27 & 3.0.1\_2018-05-15 & 4.0.1\_2019-07-19 & 4.7.7\_2025-04-19 \\
        &
        Termd & 1.0.0\_2015-10-06 & 1.1.2\_2016-08-17 & 1.1.6\_2018-10-06 & 1.1.10\_2024-11-20 \\

\midrule

\multirow{4}{*}{\shortstack{Data\\ management}} 
        &     
        JetCache & 2.1.3\_2017-04-17 & 2.5.1\_2018-05-15 & 2.6.0.RC\_2020-03-01 & 2.7.8\_2025-04-28 \\
        & 
        Gobblin & 0.12.0\_2018-06-20 & 0.13.0\_2018-09-06 & 0.15.0\_2020-11-30 & 0.17.0\_2023-06-13 \\
        & 
        Data Transfer Project & 0.1.5\_2019-02-06 & 0.3.65\_2022-06-13 & 0.4.3\_2022-12-14 & 1.1.13\_2025-05-19 \\
        & 
        P6Spy & 2.0.0\_2014-03-04 & 2.3.0\_2016-05-11 & 3.6.0\_2017-11-12 & 3.9.1\_2020-07-26hb \\

\midrule

\multirow{3}{*}{Emulator} 
        & 
        Coffee GB & 1.0.0\_2017-12-22 & 1.0.2\_2023-03-05 & 1.2.0\_2024-02-26 & 1.2.1\_2024-02-26 \\
        & 
        ZX-Poly & 2.0.0\_2019-11-20 & 2.0.7\_2020-08-16 & 2.2.3\_2022-01-15 & 2.3.5-S\_2024-12-08 \\
        & 
        unidbg & 0.3.0\_2020-06-05 & 0.8.0\_2020-11-06 & 0.9.4\_2021-07-29 & 0.9.8\_2024-09-05 \\

\midrule

\multirow{4}{*}{\shortstack{Image\\ processing}} 
        & 
        Thumbnailator & 0.4.1\_2012-04-01 & 0.4.7\_2013-12-24 & 0.4.13\_2020-10-18 & 0.4.20\_2023-06-28 \\
        & 
        ImageJ & 1.48c\_2013-09-13 & 1.51a\_2016-05-13 & 1.52s\_2019-12-12 & 1.54p\_2025-02-18 \\
        & 
        ImgLib2 & 2.0.0\_2014-10-18 & 3.3.0\_2017-02-18 & 5.9.3\_2020-07-19 & 7.1.5\_2025-04-21 \\
        & 
        Pixelitor & 4.0.0\_2016-02-18 & 4.2.1\_2019-09-26 & 4.2.4-beta2\_2021-03-13 & 4.3.1\_2023-09-06 \\

\midrule

\multirow{3}{*}{\shortstack{LLM\\ integration}} 
        & 
        LangChain Java & 0.1.5\_2023-06-07 & 0.1.8\_2023-07-06 & 0.1.11\_2023-08-04 & 0.2.2\_2023-12-21 \\
        & 
        LangChain4J & 0.4.0\_2023-06-20 & 0.18.0\_2023-07-26 & 0.30.0\_2024-01-19 & 1.0.1\_2024-04-29 \\
        & 
        Langtorch & 0.0.4\_2023-04-13 & 0.0.7\_2023-05-12 & 0.0.12\_2023-05-30 & 0.0.17\_2023-06-22 \\

\midrule

\multirow{3}{*}{\shortstack{Media\\ player}} 
        & 
        DJ Native Swing & 0.9.8\_2009-03-24 & 1.0.1\_2011-03-18 & 1.0.2\_2013-11-03 & 1.0.3\_2023-03-06 \\
        & 
        MediaPlayer & v30SepBeta\_2015-09-30 & v20151020\_2015-10-20 & v20180320\_2018-03-20 & v0.0.9.5\_2019-05-20 \\ 
        & 
        vlcj & 2.1.0\_2012-05-20 & 3.10.1\_2015-12-30 & 4.5.1.1\_2020-04-18 & 4.11.0\_2025-05-04 \\

\midrule

\multirow{3}{*}{\shortstack{Reverse\\ engineering}} 
        & 
        Recaf & 0.2\_2017-08-13 & 1.15.7\_2019-07-11 & 2.8.0\_2020-08-17 & 2.21.14\_2024-04-27 \\
        & 
        JByteMod-Beta & pre-1\_2017-09-19 & 1.4.1\_2018-01-09 & 1.6.1\_2018-03-02 & 1.8.2\_2018-08-29 \\
        & 
        JVM Explorer & v0.5.0\_2022-07-16 & v0.5.4\_2022-07-17 & v0.8.0\_2022-08-04 & v1.0.1\_2022-08-14 \\

\midrule

\multirow{3}{*}{\shortstack{Terminal\\ application}} 
        & 
        Jmxterm & 1.0.0\_2017-07-28 & 1.0.2\_2020-09-04 & 1.0.3\_2022-08-06 & 1.0.4\_2022-10-27 \\
        & 
        Muon (Snowflake) & v1.0\_2019-10-04 & v1.0.2\_2019-10-25 & v1.0.3\_2019-11-14 & v1.0.4\_2020-02-07 \\
        & 
        TN5250J & 0.6.2\_2010-12-09 & 0.7.6\_2015-10-20 & 0.8.0-beta1\_2021-03-25 & 0.8.0-beta2 \_2021-04-24 \\

\midrule

\multirow{2}{*}{Text editor} 
        & 
        RSyntaxTextArea & 1.4.1\_2010-05-01 & 2.5.8\_2015-09-09 & 3.1.5\_2021-12-30 & 3.6.0\_2025-03-18 \\
        &
        neoeedit & v353\_2021-09-19 & v378\_2022-06-12 & v383\_2023-03-01 & v387\_2024-05-22 \\

\midrule

\multirow{5}{*}{\shortstack{Unit testing\\ framework}} 
        & 
        JUnit 4 & 3.8.1\_2002-09-04 & 4.6\_2009-04-13 & 4.12-beta-2\_2014-09-25 & 4.13.2\_2021-02-13 \\
        & 
        Mockito & 1.3\_2008-04-10 & 2.0.97-beta\_2016-08-04 & 2.13.0\_2017-12-06 & 5.18.0\_2025-05-21 \\
        & 
        PowerMock & 1.4.6\_2010-10-13 & 1.5.5\_2014-05-28 & 1.7.1\_2017-08-12 & 2.0.9\_2020-11-01 \\
        & 
        Robolectric & 2.0-rc1\_2013-05-08 & 3.3.2\_2017-03-28 & 4.5.1\_2021-01-31 & 4.14.1\_2024-11-20 \\
        & 
        TestNG & 5.0\_2006-07-24 & 6.4\_2012-02-12 & 6.13\_2017-11-27 & 7.11.0\_2025-02-13 \\

\bottomrule
\end{tabular}
\end{table*}

\section{Biography Section}
\vspace{11pt}
\textbf{Nikolay Fedorov} received the B.E. degree in software engineering in 2021 and M.E. degree in computer science in 2023 from Dubna State University, Russia. 
He is currently a doctoral student of Graduate School of Environmental, Life, Natural Science and Technology, Okayama University, Japan.
His research interests include software analytics and security, as well as internet of things (IoT) systems. 

\begin{wrapfigure}{1}{25mm}
\includegraphics[width=1in,height=1.25in,clip,keepaspectratio]{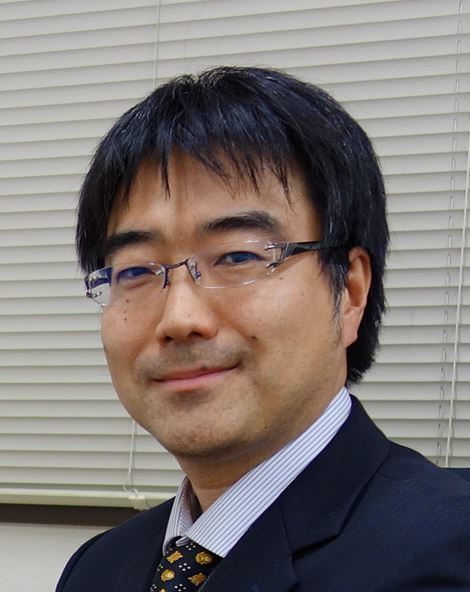}
\end{wrapfigure}\par
\noindent\textbf{Akito Monden} received the B.E. degree in electrical engineering from Nagoya University, in 1994, and the M.E. and D.E. degrees in information science from Nara Institute of Science and Technology (NAIST), in 1996 and 1998, respectively. 
He is currently a Professor with the Faculty of Environmental, Life, Natural Science and Technology, Okayama University, Japan. 
His research interests include software measurement and analytics, and software security and protection. 
He is a member of IEEE, IEICE, IPSJ, and JSSST.

\begin{wrapfigure}{1}{1in}
\includegraphics[width=1in,height=1.25in,clip,keepaspectratio]{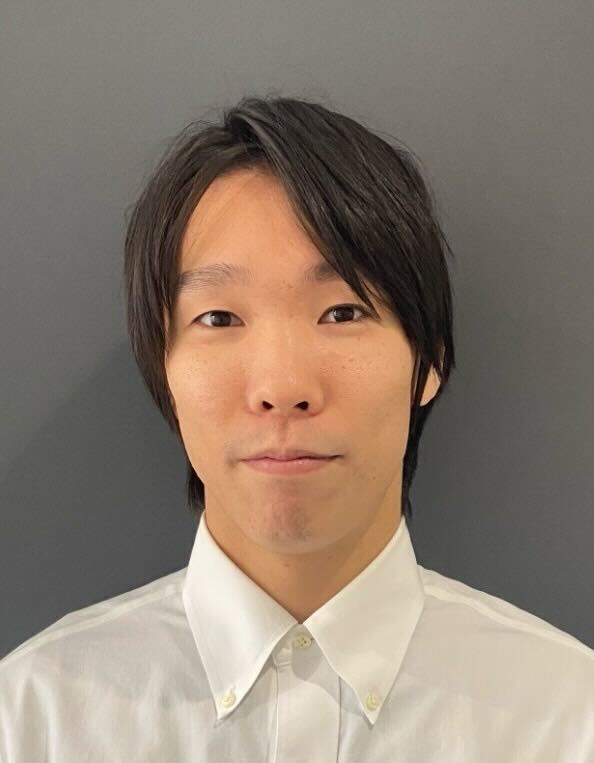}
\end{wrapfigure}\par
\noindent\textbf{Hiroki Inayoshi} received the B.E., M.E., and D.E. degrees from Nagoya Institute of Technology, Japan, in 2018, 2021, and 2024 respectively.
He is currently an Assistant Professor with the Faculty of Environmental, Life, Natural Science and Technology, Okayama University, Japan.
His research interests are in dynamic program analysis, mobile security, and privacy.
He is a member of ACM, IEEE, and IPSJ.
\begin{wrapfigure}{1}{25mm}
\includegraphics[width=1in,height=1.25in,clip,keepaspectratio]{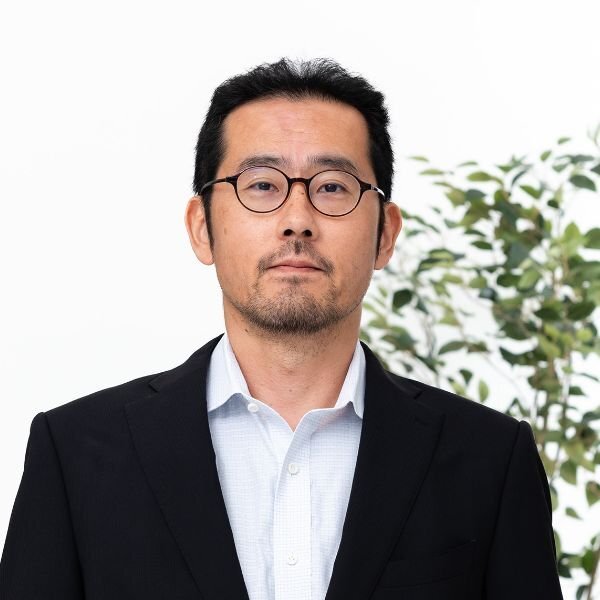}
\end{wrapfigure}\par
\noindent\textbf{Haruaki Tamada} is a professor in the Faculty of Information Science and Engineering at Kyoto Sangyo University, having joined in 2008. 
His research interests include software security and programming education. 
He received a Doctor of Engineering in Information Science from Nara Institute of Science and Technology. 
He is a member of IEICE and JPSJ.

\begin{wrapfigure}{1}{25mm}
\includegraphics[width=1in,height=1.25in,clip,keepaspectratio]{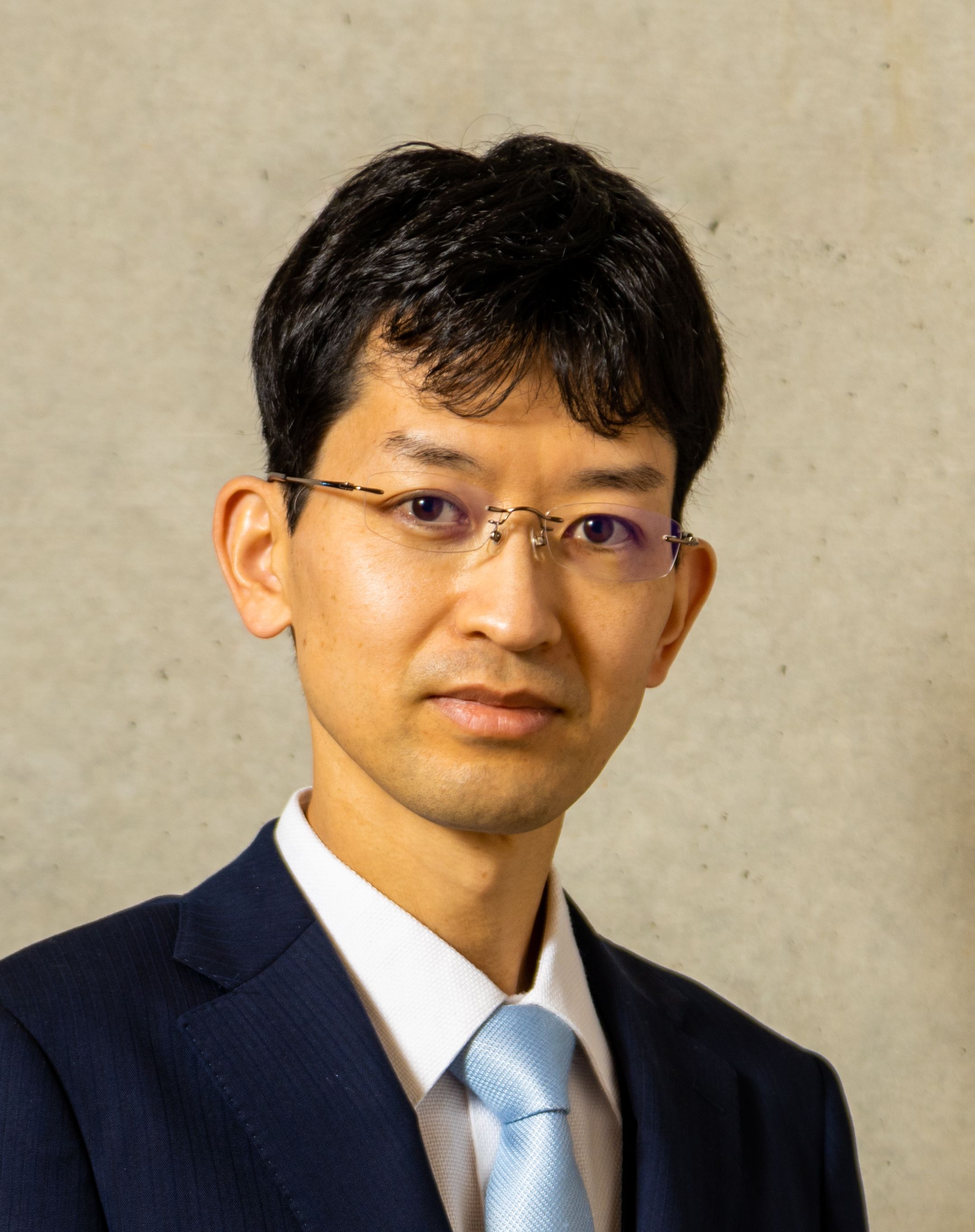}
\end{wrapfigure}\par
\noindent\textbf{Masateru Tsunoda} is an associate professor in the Department of Informatics at Kindai University, Japan. 
His research interests include software measurement and human factors in software development. 
He received a Doctor of Engineering in information science from Nara Institute of Science and Technology. 
He is a member of IEEE, IEICE, IPSJ, JSSST, and JSISE.

\end{multicols*}
\end{document}